\documentclass{article}

\usepackage[preprint]{neurips_2026}

\usepackage[utf8]{inputenc} 
\usepackage[T1]{fontenc}    
\usepackage{hyperref}       
\usepackage{url}            
\usepackage{booktabs}       
\usepackage{amsfonts}       
\usepackage{nicefrac}       
\usepackage{microtype}      
\usepackage{xcolor}         
\usepackage{graphicx}

\usepackage{subcaption}
\usepackage{amsmath}
\usepackage{multirow}
\usepackage{wrapfig}
\usepackage{tcolorbox}

\usepackage{amsthm}

\usepackage{pgfplots}
\pgfplotsset{compat=1.18}  

\usepackage{booktabs}
\usepackage{multirow}
\usepackage{graphicx}
\usepackage[table]{xcolor}

\usepackage{tikz}
\usepackage{pgfplots}
\pgfplotsset{compat=1.18} 

\newtheorem{proposition}{Proposition}

\title{WaterMoE: Expert-Routing-based Watermarking for High Fidelity and Efficiency}

\author{
Zewen Sun\\
Shanghai Jiao Tong University\\
Shanghai Innovation Institute\\
\texttt{zwsun@sjtu.edu.cn}
\And
Qian Jiang\\
Tianjin University\\
\texttt{qianjiang@tju.edu.cn}
\And
Siyuan Sheng\\
Shanghai Jiao Tong University\\
\texttt{sebestian\_1@sjtu.edu.cn}
\AND
Liyao Xiang\thanks{Corresponding author.}\\
Shanghai Jiao Tong University\\
Shanghai Innovation Institute\\
\texttt{xiangliyao08@sjtu.edu.cn}
}

\begin{document}

\maketitle

\begin{abstract}
Large language models (LLMs) have achieved remarkable success but raise growing concerns about content provenance and misuse, motivating the need for reliable watermarking techniques. However, these techniques have rarely been adopted in practice mainly for two reasons: i) severely degraded model performance, and ii) additional inference overhead. To confirm the problem, we construct a comprehensive benchmark spanning different generation tasks to systematically evaluate 9 representative watermarking methods. We found almost all existing methods are designed for text fluency, but not for restricted and complicated tasks, and their overhead prevents them from deployment in latency-critical systems.

To address i) and ii), we propose an LLM watermarking scheme \textit{WaterMoE} for the growingly popular Mixture-of-Experts (MoE) LLMs. WaterMoE embeds watermarking signals through controlled perturbation into the expert selection at each router, which accumulates to token selection shift at the final output. In contrast to watermarking as a post-processing token-sampling approach, WaterMoE embeds watermark within the inference loop incurring negligible quality degradation and computational overhead. Extensive experiments demonstrate that our method achieves a fidelity performance close to the unwatermarked and consistently outperforms state-of-the-art watermarking methods on the benchmark, with up to $4\times$ speedup, incurring merely 1\% additional inference latency compared to native generation. The results demonstrate the capability of WaterMoE to be deployed in real-world tasks.


\end{abstract}

\section{Introduction}

In recent years, Large Language Models (LLMs) have achieved remarkable success across a wide range of applications, including medical consultation   \cite{wang2025survey, yuan2024continued}, information retrieval \cite{labruna2025retrieve, zhu2025large, zeng2025rethinking, ghodratnama2023adapting},  code generation \cite{li2022competition, wang2023review, yang2025empirical, jiang2026survey} and autonomous agents \cite{huang2024understanding, liu2023dynamic, chu2025llm, li2026agentswift}. However, alongside their growing capabilities, the potential misuse of LLMs for generating misinformation, deceptive content, and malicious text has raised serious concerns about the integrity of the ecosystem \cite{liang2026watermarking, huang2026rlspoofer}. This has led to an increasing demand for reliable techniques to distinguish machine-generated text from human-written content, among which LLM watermarking has emerged as a promising direction.

\begin{wrapfigure}{r}{0.4\textwidth} 
    \centering
    \includegraphics[width=\linewidth]{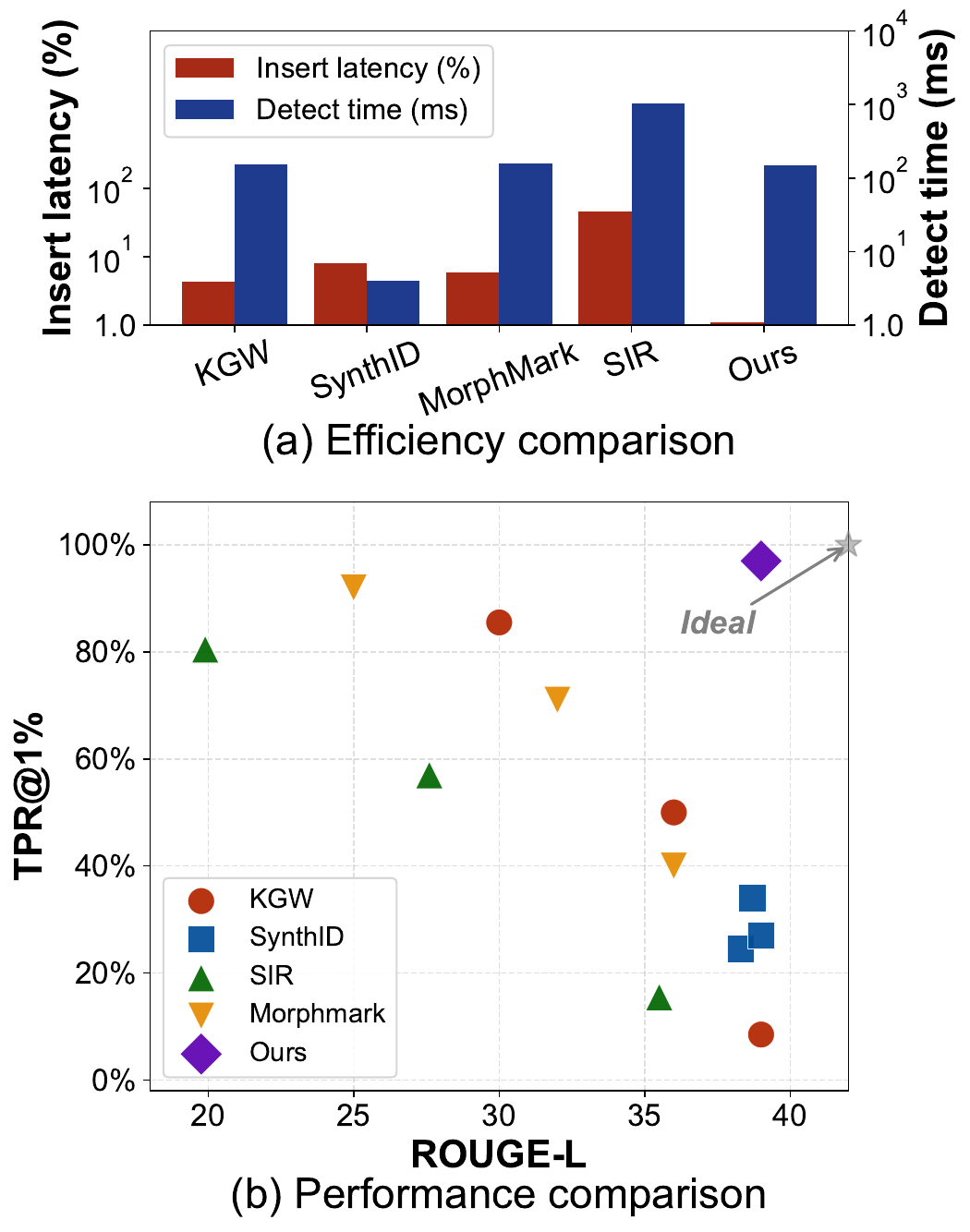}
    \caption{
    (a) Efficiency comparison on Mixtral-8x7B between WaterMoE and baseline watermarking algorithms (KGW, SynthID, SIR, and Morphmark). We report per-token insertion latency overhead (\%) and average detection time per sample (ms). (b) Performance comparison of WaterMoE  across different methods. The x-axis denotes text generation quality measured by ROUGE-L, while the y-axis represents watermark detectability (TPR@1\%). The trade-offs are obtained by varying the watermark strength. WaterMoE achieves the optimal tradeoff between detection rate and text quality.
    }
    \label{fig:multinews}
\end{wrapfigure}

Existing LLM watermarking techniques largely follow a common token-sampling-based paradigm where the model’s next-token distribution is subtly manipulated to encode identifiable patterns. For examples, KGW \cite{kirchenbauer2023watermark} partitions the vocabulary into disjoint subsets by random hashing and increases the logits of tokens in a designated subset; SynthID-text \cite{dathathri2024scalable} adopts a tournament sampling strategy to inject watermark signals without introducing distributional distortion. While being effective for post-hoc detection, these methods inherently introduce overhead in the critical decoding path, as illustrated in Fig.~\ref{fig:multinews}(a). Modern LLM serving systems are optimized for high-throughput, low-latency generation, where even lightweight per-token operations---such as hash-based partitioning, probability reweighing, or additional sampling---accumulate linearly with sequence length, leading to increased latency and significantly reduced system efficiency.
At the same time, these approaches rely on explicitly altering token distributions, which inevitably interferes with the model’s natural generation behavior. This interference is particularly detrimental in delicate real-world tasks such as summarization, instruction following, code generation, etc. In those restricted settings, valid outputs are sparse in the token space and thus token probability shifting caused by watermarking tends to drive the generated text to deviate from correct or optimal outputs. As shown in Fig.~\ref{fig:multinews}(b), we evaluate existing watermarking methods under different parameter settings on MultiNews for a summarization task. The results show a clear `impossible' region at the top-right corner indicating high detection rates and high fidelity.


Therefore, a fundamental question arises: \emph{can we design a watermarking mechanism that achieves high efficiency, fidelity and detection rate at the same time?}

To answer the question, we propose \textit{WaterMoE}, an expert-routing-based watermarking approach for Mixture-of-Experts (MoE) LLMs. MoE models have been increasingly adopted in modern high-performance language models \cite{cai2024survey, zhu2024llama}, where inference is carried out through sparse expert activation and routing decisions. We are motivated by the inherent redundancy in MoE models that allows different experts to be chosen so that each input token takes a flexible combination of model weights, leading to a varied output distribution \cite{rajbhandari2022deepspeed, skliar2024mixture}. Hence without altering the model’s generation strategy, the expert selection brings in `controlled randomness' in the output distribution: at each routing gate, top-$k$ experts are potentially chosen, allowing stable and detectable internal patterns to be expressed through subtle biases at the routing level, while maintaining generation capability. Building on this observation, we integrate watermark signals directly into the MoE routing logic, making the watermark an intrinsic component of the model’s inference process.



Our contributions are summarized as follows.
1) \textbf{Real-world evaluation of watermarking}:
we systematically reveal the limitations of existing watermarking methods under realistic workloads, and construct a comprehensive benchmark spanning summarization, question answering, code generation, reasoning tasks and instruction following to evaluate the watermark detection rate, efficiency, and task performance.
2) \textbf{Structural watermarking via MoE routing}: we propose \textbf{WaterMoE}, a watermarking approach that embeds signals into expert routing.
Extensive experiments show that WaterMoE improves detection under strict false positive constraints by 12.1\% at TPR@1\% over the strongest baseline, without degradation from unwatermarked task performance. Our method achieves up to 4$\times$ speedup in watermark embedding compared to state-of-the-art methods, incurring merely 1\% latency overhead compared to native generation, demonstrating strong practicality.

\section{Preliminary and Related Work}


In recent years, Mixture-of-Experts (MoE) has been increasingly adopted in production-level large language models for online services, such as DeepSeek-V3 \cite{liu2024deepseek} and Qwen3 \cite{yang2025qwen3}.
Mixture-of-Experts (MoE) architectures introduce a routing mechanism that enables sparse activation of model parameters during inference \cite{cai2024survey, rajbhandari2022deepspeed}. In a typical MoE layer, the standard feed-forward network is replaced by multiple parallel expert networks, while a router dynamically selects a small subset of experts for each token based on routing scores. Only the top-$k$ experts are activated, and their outputs are aggregated through a weighted sum gate \cite{li2025uni}. This design allows MoE models to scale up model capacity while keeping inference computationally bounded, which has been widely adopted in modern large-scale language models.

Formally, given the hidden representation $\mathbf{h}_\ell$ at layer $\ell$, the router produces routing scores $\mathbf{s}_\ell \in \mathbb{R}^K$, where $K$ denotes the number of experts in layer $\ell$. The scores are converted into a routing distribution via a softmax, the top-$k$ of which are selected as experts $\mathcal{E}_\ell(\mathbf{h}_\ell)$ to compute the output of the layer $\mathrm{MoE}_\ell(\mathbf{h}_\ell)$:
\begin{equation*}
\mathbf{s}_\ell = \mathrm{Router}_\ell(\mathbf{h}_\ell),~~
p_\ell(i \mid \mathbf{h}_\ell) = \mathrm{softmax}(\mathbf{s}_\ell)_i, ~~~ \mathrm{MoE}_\ell(\mathbf{h}_\ell) = \sum_{i \in \mathcal{E}_\ell(\mathbf{h}_\ell)} p_\ell(i \mid \mathbf{h}_\ell)\, f_i(\mathbf{h}_\ell),
\end{equation*}
where $f_i(\cdot)$ denotes the $i$-th expert network and $p_\ell(i \mid \mathbf{h}_\ell)$ is the corresponding routing weight. More details of the MoE framework are provided in Appendix \ref{appendix_moe}.

\textbf{LLM watermarking} methods mostly embed watermark signals by modifying the token logits distribution in autoregressive decoding. KGW \cite{kirchenbauer2023watermark} partitions the vocabulary via N-gram hashing and biased generation toward a selected subset. EWD \cite{lu2024entropy} improves detection rate by entropy-aware embedding, SynthID \cite{dathathri2024scalable} proposes a non-distortionary watermarking methods through tournament sampling, and MorphMark \cite{wang2025morphmark} takes an adaptive watermarking approach.
However, most token-sampling-based methods rely on post-hoc logit perturbations, causing text quality degradation and additional per-token generation overhead (e.g., hashing, dynamic reweighing, etc.) that are difficult to mitigate. The quality degradation is attributed to the conflict between the sampling constraints posed by watermarking and the harsh requirement in real-world generation including factual correctness, strict formatting, program executability, etc.: the former often asks for a deviation from the original output while the latter forbids that. Hence the current approach often has to sacrifice watermarking accuracy for text quality (as shown in Fig.~\ref{fig:multinews}). The additional overhead largely stems from the need for reproducible randomness, typically implemented via per-token hashing, which breaks the fully GPU-resident execution pipeline, leading to noticeable inference latency. While precomputing randomness on GPU could in principle remove this cost, it would require non-trivial GPU kernel engineering or introduce additional memory and implementation overhead.

In contrast, our method embeds watermark signals into the expert routing mechanism of Mixture-of-Experts models. By introducing perturbation in the expert selection rather than the token choice, watermark embedding is seamlessly integrated into routing priors, making it an inherent part of inference thus incurring negligible latency overhead while preserving output quality to a large extent.

\section{WaterMoE: Watermarking via Expert Routing}

\begin{figure}[t]
    \centering
    \includegraphics[width=\linewidth]{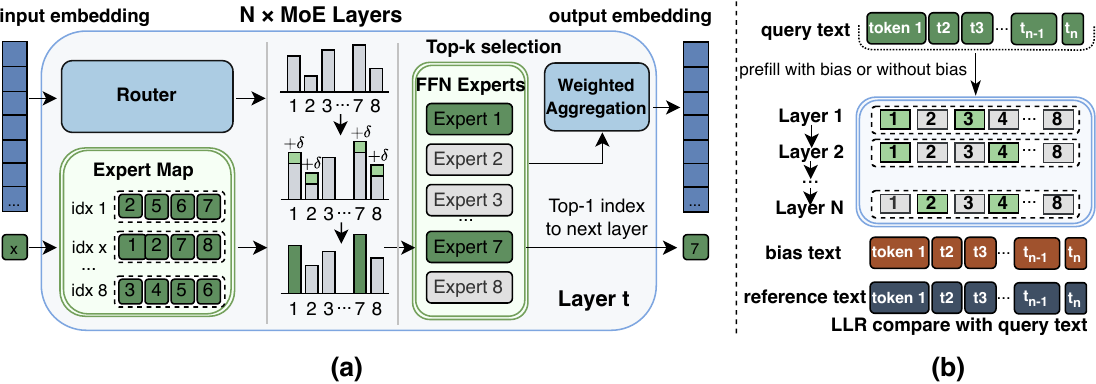}
    \caption{
       Overview of the WaterMoE framework.
{(a) Watermark embedding via routing bias.}
A lightweight additive bias is applied to the routing scores of green experts, subtly steering top-$k$ expert selection.
{(b) Watermark detection via likelihood test.}
The presence of watermark signals is identified by a token-level likelihood test between biased and reference models, where consistent shifts reveal statistically detectable watermark patterns.
    }
    \label{fig:MoE}
\end{figure}

Our design is motivated by the flexibility in MoE models' routing mechanism: experts with similar routing scores often possess comparable functional capabilities. Hence a small routing bias can shift expert selection which gradually amplifies through layers leading to subtle, but traceable variations in the final output distribution. Since experts with similar capabilities are selected, our framework preserves fidelity while being highly efficient. The proposed framework is shown in Fig.~\ref{fig:MoE}.

\subsection{Watermark Embedding through Routing Bias}
\label{sec:expert_map}




To enable efficient watermark embedding at scale, we introduce a \emph{Green Expert Map} $\mathcal{M}_\ell$ that is randomly generated offline, and deterministically specifies the green experts for each MoE layer. The green experts refer to those experts perturbed by the injected biases.
Let $e_{\ell-1} \in \{1, \dots, K\}$ denote the top-1 selected expert index from layer $\ell-1$. The green expert set at layer $\ell$ is then determined as:
\begin{equation}
\mathcal{G}_\ell = \mathcal{M}_\ell(e_{\ell-1}), \quad \text{s.t.} \quad \mathcal{G}_\ell \subseteq \{1, \dots, K\}, \quad |\mathcal{G}_\ell| = K/2. 
\end{equation}
For the first MoE layer, we skip this procedure due to the absence of a preceding expert. We utilize a binary indicator vector $\mathbf{g}_\ell \in \{0,1\}^K$ to denote the selected green set:
\begin{equation}
g_{\ell,i} =
\begin{cases}
1, & i \in \mathcal{G}_\ell, \\
0, & \text{otherwise}.
\end{cases}
\end{equation}
The routing bias for layer $\ell$ is defined by $\delta \, \mathbf{g}_\ell$, where $\delta > 0$ signaling the watermark strength. Hence the biased routing logits are:
\begin{equation}
\tilde{\mathbf{s}}_\ell = \mathbf{s}_\ell + \delta \, \mathbf{g}_\ell.
\label{eq:routingbias}
\end{equation}
We further define the corresponding routing probabilities $\mathbf{p}_{\ell}$ and $\tilde{\mathbf{p}}_{\ell}$ for $\mathbf{s}_\ell$ and $\tilde{\mathbf{s}}_\ell$, respectively, by passing them through a Softmax function.
\begin{proposition}[First-order invariance of routing distribution]
Let $p_{\ell,i}$ and $\tilde{p}_{\ell,i}$ denote the routing probabilities of the $i$-th expert in layer $\ell$ before and after bias injection. Then the expected routing probability over the random choice of $i \in \mathcal{G}_\ell$ satisfies
\[
\mathbb{E}_{i}\big[\tilde{p}_{\ell,i}\big]
= p_{\ell,i} + \mathcal{O}(\delta^2).
\]
\end{proposition}
A rigorous proof of this property is provided in Appendix \ref{appendix_prove}. The property indicates that with the first-order term vanishing, the routing distribution shifts by at most the second order of $\delta$, which is almost negligible. 

In practice, the green expert map is precomputed offline and stored as a tensor aligned with MoE layers. During inference, the routing bias is applied via a single element-wise addition (Eq.~\eqref{eq:routingbias}), which is fully GPU-parallelizable and introduces negligible latency overhead.

During autoregressive generation, the routing bias is consistently applied to every token across all MoE layers. As a result, the expert selection process is subtly biased toward green experts. Although the token logits are not explicitly modified, the perturbed routing alters the internal representations that ultimately induce subtle but consistent shifts in the output distribution, enabling statistical detection using standard hypothesis testing methods (e.g., Z-test).


\subsection{Likelihood-Based Watermark Detection via Reference Calibration}
\label{sec:detection}
For watermark detection, the key idea is to detect the distributional shift caused by routing biases. We use an unwatermarked model to estimate the intrinsic likelihood of each token as a reference. Then we calibrate token-level probabilities through disentangling routing-induced preference from the native generations, thereby isolating the watermark signals.

Given a sequence $\mathbf{x} = (x_1, \dots, x_T)$, we evaluate it using a \emph{single model} under two routing configurations: i) a biased mode with the watermarking enabled, and ii) a reference mode without routing bias. Importantly, both configurations share the same model parameters and differ only in the expert routing path, thus needing no two separate model deployments.

For each token, we compute the log likelihood difference as:
\begin{equation}
\Delta_t = \ell_t^{\mathrm{bias}} - \ell_t^{\mathrm{base}}, \quad
\ell_t^{\mathrm{bias}} = \log P_{\mathrm{bias}}(x_t \mid x_{<t}), \quad
\ell_t^{\mathrm{base}} = \log P_{\mathrm{base}}(x_t \mid x_{<t}),
\end{equation}
where $P_{\mathrm{bias}}$ and $P_{\mathrm{base}}$ denote the conditional distributions of $x_t$ under configuration i) and ii). A higher $\ell_t^{\mathrm{base}}$ indicates that the token is highly consistent with the context. $\Delta_t$ measures the additional preference induced by the routing bias for the token. 
We aggregate token-level signals of which the log likelihood difference exceeds threshold $\tau$:
\begin{equation}
S(\mathbf{x}; \tau) = \frac{1}{T} \sum_{t=1}^T \mathbb{I}(\Delta_t > \tau).
\end{equation}

On the reference model, we estimate the expectation $\mu_0(\tau)$ and standard deviation $\sigma_0(\tau)$ of $S(\mathbf{x}; \tau)$, and calibrate it by:
\begin{equation}
Z(\mathbf{x}) = \frac{S(\mathbf{x}; \tau) - \mu_0(\tau)}{\sigma_0(\tau)}.
\end{equation}
A sequence is flagged as watermarked if $Z(\mathbf{x})$ exceeds a threshold.

\textbf{Why works?} WaterMoE achieves efficiency by significantly reducing the watermark embedding time at the cost of mildly increased detection overhead. Previous LLM watermarking methods contain inefficient hashing in both watermark embedding and detection for reproducible randomness given the context. However, embedding is far more frequent than detection in practice: every token generated is required to be watermarked but detection is upon request. Thus we choose to remove hashing but introduce perturbed expert mapping in the embedding. Such expert mapping is on one hand of small scale, compared with the perturbation at the logits; on the other, the perturbation gets progressively amplified through the model’s feed-forward dynamics, resulting in consistent and detectable statistical signals in the output distribution without relying on explicit hashing.

WaterMoE also produces output of high quality. The reason is that the routing biases primarily influence the selection among functionally similar experts, ensuring that the model’s internal representations are only mildly affected. This allows the model to retain sufficient flexibility to explore the full token space and to self-correct through subsequent forward passes, thereby preserving performance even in highly restricted scenarios. As these subtle routing shifts accumulate, they induce lightweight yet consistent deviations at the token level: the generated outputs remain valid and task-compliant, while exhibiting slight statistical differences from the original distribution, which can be reliably exploited for watermark detection.

\section{Experimental Results}

To comprehensively evaluate the effectiveness of WaterMoE, we organize our experiments around the following research questions.
    \textit{RQ1 Detection Performance:} 
    Can WaterMoE achieve reliable and highly accurate detection across different models and tasks?
    \textit{RQ2 Generation Quality:} How does WaterMoE affect the generation quality?
    \textit{RQ3 Efficiency:} 
    What is the computational overhead of watermark embedding and detection compared to existing methods?
    \textit{RQ4 Robustness:} 
    How robust is WaterMoE under realistic adversarial attacks?
    \textit{RQ5 Stealthiness:} 
    Can WaterMoE remain statistically indistinguishable from unwatermarked text under black-box detection?

\subsection{Setup}
All experiments are implemented based on the MarkLLM \cite{pan2024markllm} repository and conducted on NVIDIA H200 GPUs.
We consider two representative MoE models:
Mixtral-8$\times$7B-Instruct-v0.1 \cite{jiang2024mixtral}and Qwen3-30B-A3B-Instruct \cite{yang2025qwen3}, both of which adopt sparse expert routing mechanisms. We compare WaterMoE against a set of representative watermarking baselines, as summarized in Table \ref{tab:high_complexity_results}. For WaterMoE, we set as default $\delta = 0.2$ and $\tau = 0$ across all experiments. The complete experimental setup and hyperparameter configurations are provided in Appendix~\ref{appendix_watermark_setup}.

We construct a benchmark suite spanning diverse LLM evaluation settings, organized into three tiers by task complexity, as illustrated in Table~\ref{tab:dataset_overview}: \textit{Low-}, \textit{Moderate-}, and \textit{High-Complexity} tasks. This tiered design enables a systematic assessment of watermark performance under varying difficulty and constraint levels. Notably, the low-complexity setting where $\delta = 0.8$ is included to ensure fair comparison with existing watermarking methods.

\begin{table}[htbp]
  \caption{Overview of evaluated datasets grouped by task complexity.‘Len.(In/Ans)’ refer to the average length of input question and reference answer.}
  \label{tab:dataset_overview}
  \centering
  \resizebox{\linewidth}{!}{
  \begin{tabular}{llclllc}
    \toprule
    \textbf{Complexity} & \textbf{Source Data} & \textbf{ID} & \textbf{Task} & \textbf{Metric} & \textbf{Language} & \textbf{Len. (In/Ans)} \\
    \midrule
    
    \multirow{3}{*}{\textbf{Low}} & \multicolumn{6}{l}{\textit{(Text Gen.)}} \\
    & \quad C4 \cite{raffel2020exploring} & 1-1 & Language Modeling & PPL & English & 22.14 / 151.01 \\
    & \quad Booksum \cite{kryscinski2022booksum} & 1-2 & Long-Context Generation & PPL & English & 45.53 / 541.54 \\
    \midrule

    \multirow{3}{*}{\textbf{Moderate}} & \multicolumn{6}{l}{\textit{(Sum. and QA)}} \\
    & \quad Multinews \cite{fabbri2019multi} & 2-1 & Multi-Doc Sum. & Rouge-L & English & 1464.55 / 220.50 \\
    & \quad ELI5 \cite{fan2019eli5} & 2-2 & Long-Form QA & Rouge-L & English & 38.98 / 261.31 \\
    \midrule

    \multirow{9}{*}{\textbf{High}} & \multicolumn{6}{l}{\textit{(Code Gen.)}} \\
    & \quad APPS \cite{hendrycks2021measuring} & 3-1 & Competitive Coding & PASS@1 & Python & 313.52 / 108.80 \\
    & \quad CodeContests \cite{li2022competition} & 3-2 & Competitive Coding & PASS@1 & Python/C\#/Java & 383.45 / 176.75 \\
    \cmidrule{2-7} 
    
    & \multicolumn{6}{l}{\textit{(CoT Reasoning)}} \\
    & \quad GSM8K \cite{cobbe2021training} & 3-3 & Math Word Problem & Accuracy & English & 54.39 / 151.66 \\
    & \quad MMLU \cite{hendrycks2020measuring} & 3-4 & Multiple Choice QA & Accuracy & English & 54.35 / 256.53 \\
    \cmidrule{2-7}
    
    & \multicolumn{6}{l}{\textit{(Alignment)}} \\
    & \quad IFEval \cite{zhou2023instruction} & 3-5 & Instruction Following & Rule-based Judge & English & 37.42 / 1502.53 \\
    & \quad WritingBench \cite{wu2025writingbench} & 3-6 & Creative Generation & LLM-based Score & English & 482.04 / 2512.34 \\
    
    \bottomrule
  \end{tabular}
  }
\end{table}

\subsection{RQ1 \& RQ2: Detection Performance and Generation Quality}
\textbf{Performance on High-Complexity Tasks.} Table~\ref{tab:high_complexity_results} reports watermarking performance across highly constrained domains which impose rigid structural and logical requirements, making watermark injection particularly challenging. Under a strict false positive rate (TPR@1\%), baseline methods (e.g., EWD, SynthID) exhibit notable degradation in detectability, often falling below 70\% on reasoning and coding benchmarks such as GSM8K, MMLU, and CodeContests.
EXPEdit performs poorly in such settings, as it operates at the sample level, where the space of valid edits is severely constrained in high-complexity tasks, making watermark injection difficult.
In contrast, WaterMoE achieves consistently strong detection performance, exceeding 90\% on challenging datasets like APPS and CodeContests, and reaching 100.0\% on alignment benchmarks (IFEval, WritingBench). Importantly, this high detectability does not come at the cost of generation quality: while conventional methods (e.g., KGW, SynthID) incur substantial accuracy drops (over 15 points) on strict tasks, WaterMoE maintains performance closely aligned with the unwatermarked model across all datasets, effectively mitigating the detectability–quality trade-off under rigid constraints. The results for moderate-complexity tasks are provided in Appendix \ref{appendix_modcom}.

\begin{table}[htbp]
  \caption{Watermarking performance on high-complexity tasks. WaterMoE maintains near-native generation quality on these highly rigid tasks while consistently achieving state-of-the-art detectability. Here, Accuracy and PASS@1 denote the percentage (\%) of  correct answers. AUC indicates the area under the ROC curve for detection performance.}
  \label{tab:high_complexity_results}
  \centering
  \resizebox{\linewidth}{!}{
  \begin{tabular}{lcccccccc}
    \toprule
    
    \multicolumn{9}{c}{\textbf{Category: Code Generation (Model: Qwen)}} \\
    \midrule
    \multirow{2}{*}{\textbf{Algorithm}} & \multicolumn{4}{c}{\textbf{Dataset: APPS}} & \multicolumn{4}{c}{\textbf{Dataset: CodeContests}} \\
    \cmidrule(lr){2-5} \cmidrule(lr){6-9}
    & \textbf{TPR@1\%} & \textbf{TPR@5\%} & \textbf{AUC} & \textbf{PASS@1} & \textbf{TPR@1\%} & \textbf{TPR@5\%} & \textbf{AUC} & \textbf{PASS@1} \\
    \midrule
    
    \rowcolor{gray!15} \textit{Unwatermarked} & - & - & - & 65.4 & - & - & - & 42.0 \\
    EWD \cite{lu2024entropy} & 73.0 & 92.0 & 97.3 & 27.8 & 40.0 & 54.0 & 92.3 & 42.5 \\
    EXPEdit \cite{kuditipudi2023robust}& 6.0 & 6.0 & 41.3 & 2.0 & 2.0 & 2.0 & 39.9 & 6.1 \\
    KGW \cite{kirchenbauer2023watermark}& 63.0 & 72.0 & 94.0 & 42.2 & 38.0 & 42.0 & 79.4 & 34.4 \\
    MorphMark \cite{wang2025morphmark}& 71.0 & 82.0 & 96.3 & 42.5 & 46.0 & 50.0 & 83.3 & 42.2 \\
    SynthID \cite{dathathri2024scalable}& 80.0 & 87.0 & 92.0 & 41.1 & 64.0 & 78.0 & 89.2 & 40.5 \\
    Unbiased \cite{hu2023unbiased}& 53.0 & 63.0 & 86.5 & 58.9 & 41.0 & 56.0 & 82.2 & 41.4 \\
    UPV \cite{liu2023unforgeable}& 62.0 & 78.0 & 92.3 & 35.4 & 32.0 & 36.0 & 62.8 & 41.0 \\
    \midrule
    \rowcolor{gray!15} \textbf{WaterMoE} & \textbf{90.0} & \textbf{98.0} & \textbf{99.6} & \textbf{61.9} & \textbf{92.0} & \textbf{98.0} & \textbf{99.6} & \textbf{44.2} \\
    
    \midrule
    \midrule
    
    \multicolumn{9}{c}{\textbf{Category: Strict Alignment (Model: Qwen)}} \\
    \midrule
    \multirow{2}{*}{\textbf{Algorithm}} & \multicolumn{4}{c}{\textbf{Dataset: IFEval}} & \multicolumn{4}{c}{\textbf{Dataset: WritingBench}} \\
    \cmidrule(lr){2-5} \cmidrule(lr){6-9}
    & \textbf{TPR@1\%} & \textbf{TPR@5\%} & \textbf{AUC} & \textbf{Accuracy} & \textbf{TPR@1\%} & \textbf{TPR@5\%} & \textbf{AUC} & \textbf{Accuracy} \\
    \midrule
    
    \rowcolor{gray!15} \textit{Unwatermarked} & - & - & - & 88.5 & - & - & - & 91.3 \\
    EWD & 97.0 & 100.0 & 99.7 & 87.7 & 100.0 & 100.0 & 100.0 & 91.3 \\
    EXPEdit & 61.0 & 61.0 & 83.5 & 60.1 & 30.0 & 30.0 & 69.4 & 57.0 \\
    KGW & 60.0 & 80.0 & 93.0 & 87.7 & 88.0 & 98.0 & 98.7 & 90.8 \\
    MorphMark & 50.0 & 77.0 & 91.6 & 88.3 & 51.0 & 88.0 & 96.3 & 91.1 \\
    SynthID & 99.0 & 99.0 & 99.9 & 41.1 & 100.0 & 100.0 & 100.0 & 91.0 \\
    Unbiased & 80.0 & 85.0 & 95.6 & 88.3 & 54.0 & 70.0 & 83.7 & 91.1 \\
    UPV & 20.0 & 60.0 & 53.1 & 87.7 & 1.0 & 1.0 & 54.4 & 91.5 \\
    \midrule
    \rowcolor{gray!15} \textbf{WaterMoE} & \textbf{100.0} & \textbf{100.0} & \textbf{100.0} & \textbf{88.3} & \textbf{100.0} & \textbf{100.0} & \textbf{100.0} & \textbf{91.6} \\
    
    \midrule
    \midrule
    
    \multicolumn{9}{c}{\textbf{Category: Chain-of-Thought Reasoning (Model: Qwen)}} \\
    \midrule
    \multirow{2}{*}{\textbf{Algorithm}} & \multicolumn{4}{c}{\textbf{Dataset: GSM8K}} & \multicolumn{4}{c}{\textbf{Dataset: MMLU}} \\
    \cmidrule(lr){2-5} \cmidrule(lr){6-9}
    & \textbf{TPR@1\%} & \textbf{TPR@5\%} & \textbf{AUC} & \textbf{Accuracy} & \textbf{TPR@1\%} & \textbf{TPR@5\%} & \textbf{AUC} & \textbf{Accuracy} \\
    \midrule
    
    \rowcolor{gray!15} \textit{Unwatermarked} & - & - & - & 72.0 & - & - & - & 61.0 \\
    EWD & 79.0 & 83.0 & 94.8 & 53.0 & 85.0 & 93.0 & 98.9 & 54.0 \\
    EXPEdit & 11.0 & 22.0 & 67.2 & 33.0 & 3.0 & 14.0 & 42.2 & 59.0 \\
    KGW & 69.0 & 78.0 & 95.6 & 54.0 & 67.0 & 81.0 & 96.1 & 49.0 \\
    MorphMark & 44.0 & 49.0 & 85.6 & 68.0 & 43.0 & 47.0 & 84.9 & 62.0 \\
    SynthID & 70.0 & 82.0 & 93.9 & 68.0 & 82.0 & 86.0 & 94.8 & 47.0 \\
    Unbiased & 58.0 & 62.0 & 83.9 & 66.0 & 69.0 & 80.0 & 89.3 & 56.0 \\
    UPV & 58.0 & 72.0 & 93.2 & 38.0 & 61.0 & 80.0 & 95.3 & 47.0 \\
    \midrule
    \rowcolor{gray!15} \textbf{WaterMoE} & \textbf{91.0} & \textbf{95.0} & \textbf{99.2} & \textbf{68.0} & \textbf{94.0} & \textbf{96.0} & \textbf{99.6} & \textbf{63.0} \\
    
    \bottomrule
  \end{tabular}
  }
\end{table}

\textbf{Performance on Low-Complexity Tasks.}
Table~\ref{tab:low_complexity_results} presents results on low-complexity, open-ended generation tasks (C4 and Booksum). While such tasks are generally friendly to watermarking, the detection performance varies across model architectures. On Mixtral, most methods achieve near-perfect detection, but noticeable degradation occurs on Qwen under strict false positive constraints (TPR@1\%), where baselines such as MorphMark, SynthID, and KGW exhibit significant drops. In contrast, WaterMoE maintains consistent 100.0\% detection rate across all models and datasets, demonstrating strong model-independent stability. Importantly, linguistic fluency is not sacrificed: while some baselines (e.g., EXPEdit, EWD) substantially raises perplexity, WaterMoE achieves highly competitive PPLs among watermarked models, closely aligning with the unwatermarked baseline. These results indicate that WaterMoE effectively preserves fluency while ensuring reliable detectability in unconstrained generation settings.

\begin{table}[htbp]
  \caption{Watermarking performance on low-complexity tasks. Detectability metrics (TPR and AUC) are reported as percentages (higher is better), while text quality is evaluated by Perplexity (PPL, lower is better). The best results are highlighted in bold.}
  \label{tab:low_complexity_results}
  \centering
  \resizebox{\linewidth}{!}{
  \begin{tabular}{lcccccccc}
    \toprule
    
    \multicolumn{9}{c}{\textbf{Dataset: C4}} \\
    \midrule
    \multirow{2}{*}{\textbf{Algorithm}} & \multicolumn{4}{c}{\textbf{Mixtral-8x7B}} & \multicolumn{4}{c}{\textbf{Qwen}} \\
    \cmidrule(lr){2-5} \cmidrule(lr){6-9}
    & \textbf{TPR@1\%} & \textbf{TPR@5\%} & \textbf{AUC} & \textbf{PPL} $\downarrow$ 
    & \textbf{TPR@1\%} & \textbf{TPR@5\%} & \textbf{AUC} & \textbf{PPL} $\downarrow$ \\
    \midrule
    
    \rowcolor{gray!15} \textit{Unwatermarked} & - & - & - & 6.18 & - & - & - & 5.22 \\
    EWD \cite{lu2024entropy}& 100.0 & 100.0 & 100.0 & 8.16 & 94.0 & 100.0 & 99.7 & 6.76 \\
    EXPEdit \cite{kuditipudi2023robust}& 93.5 & 96.0 & 98.7 & 8.49 & 98.0 & 98.5 & 99.4 & 8.79 \\
    KGW \cite{kirchenbauer2023watermark}& 100.0 & 100.0 & 100.0 & 7.80 & 82.0 & 88.0 & 98.2 & 6.61 \\
    MorphMark \cite{wang2025morphmark}& 100.0 & 100.0 & 100.0 & 6.85 & 87.0 & 95.5 & 98.4 & 6.51 \\
    SIR \cite{liu2023semantic}& 100.0 & 100.0 & 100.0 & 7.65 & - & - & - & - \\
    SynthID \cite{dathathri2024scalable}& 100.0 & 100.0 & 100.0 & \textbf{6.12} & 86.5 & 90.0 & 96.8 & 6.44 \\
    Unbiased \cite{hu2023unbiased}& 100.0 & 100.0 & 100.0 & 6.72 & 96.0 & 99.0 & 99.9 & 6.39 \\
    UPV \cite{liu2023unforgeable}& 100.0 & 100.0 & 100.0 & 7.10 & 81.0 & 90.0 & 97.2 & 6.27 \\
    XSIR \cite{he2024can} & 99.5 & 100.0 & 99.9 & 7.68 & - & - & - & - \\
    \midrule
    \rowcolor{gray!15} \textbf{WaterMoE} & \textbf{100.0} & \textbf{100.0} & \textbf{100.0} & 6.63 
    & \textbf{100.0} & \textbf{100.0} & \textbf{100.0} & \textbf{5.48} \\
    
    \midrule
    \midrule
    
    \multicolumn{9}{c}{\textbf{Dataset: Booksum}} \\
    \midrule
    \multirow{2}{*}{\textbf{Algorithm}} & \multicolumn{4}{c}{\textbf{Mixtral-8x7B}} & \multicolumn{4}{c}{\textbf{Qwen}} \\
    \cmidrule(lr){2-5} \cmidrule(lr){6-9}
    & \textbf{TPR@1\%} & \textbf{TPR@5\%} & \textbf{AUC} & \textbf{PPL} $\downarrow$ 
    & \textbf{TPR@1\%} & \textbf{TPR@5\%} & \textbf{AUC} & \textbf{PPL} $\downarrow$ \\
    \midrule
    
    \rowcolor{gray!15} \textit{Unwatermarked} & - & - & - & 6.94 & - & - & - & 5.74 \\
    EWD & 100.0 & 100.0 & 100.0 & 9.04 & 98.0 & 98.0 & 99.8 & 7.35 \\
    EXPEdit & 99.0 & 99.0 & 99.9 & 10.50 & 100.0 & 100.0 & 99.9 & 11.31 \\
    KGW & 100.0 & 100.0 & 100.0 & 8.45 & 84.5 & 92.0 & 98.0 & 7.42 \\
    MorphMark & 100.0 & 100.0 & 100.0 & 7.86 & 89.0 & 95.5 & 98.6 & 7.30 \\
    SIR & 100.0 & 100.0 & 100.0 & 8.08 & - & - & - & - \\
    SynthID & 100.0 & 100.0 & 100.0 & \textbf{6.80} & 92.0 & 94.0 & 98.7 & 7.47 \\
    Unbiased & 100.0 & 100.0 & 100.0 & 7.56 & 93.0 & 99.0 & 99.8 & 7.42 \\
    UPV & 100.0 & 100.0 & 100.0 & 7.99 & 67.5 & 85.0 & 92.8 & 7.52 \\
    XSIR & 100.0 & 100.0 & 100.0 & 8.59 & - & - & - & - \\
    \midrule
    \rowcolor{gray!15} \textbf{WaterMoE} & \textbf{100.0} & \textbf{100.0} & \textbf{100.0} & 7.50 
    & \textbf{100.0} & \textbf{100.0} & \textbf{100.0} & \textbf{6.37} \\
    
    \bottomrule
  \end{tabular}
  }
\end{table}

\begin{figure}[t]
    \centering
    \begin{subfigure}{0.48\linewidth}
        \centering
        \includegraphics[width=\linewidth]{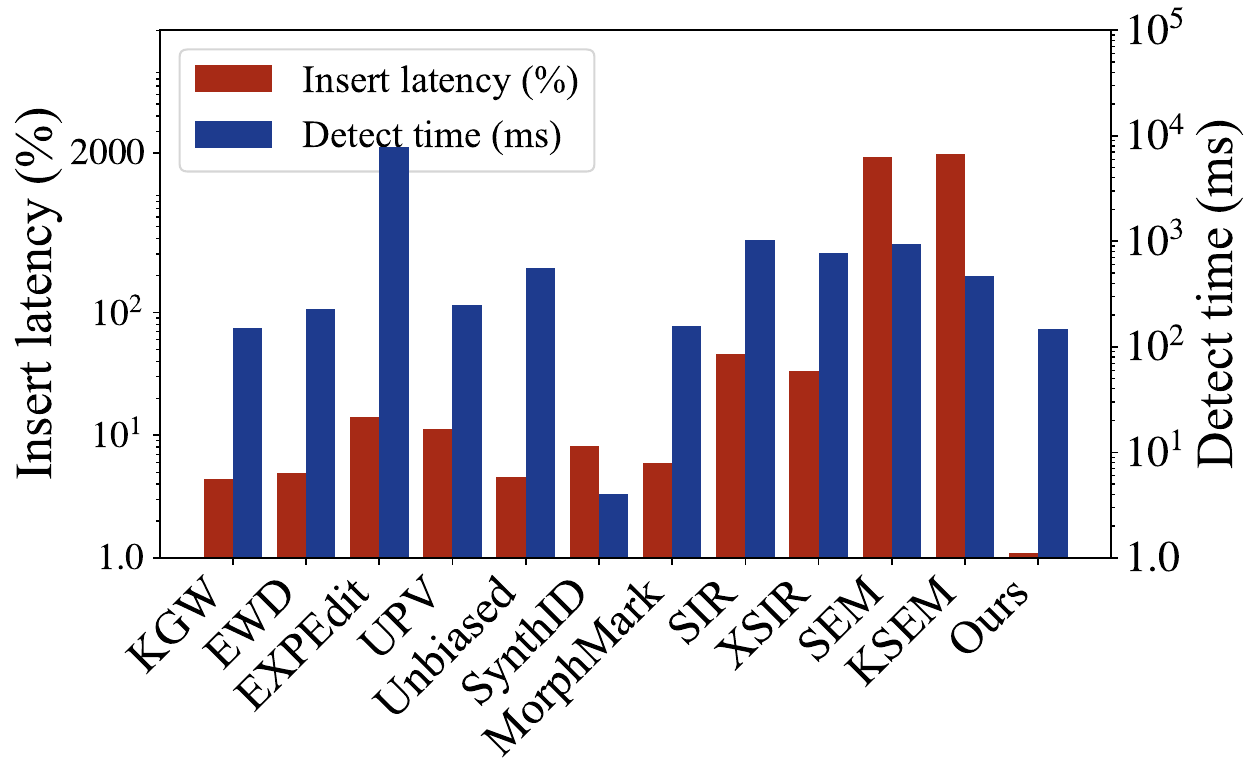}
        \caption{Efficiency comparison.}
        \label{fig:sub_efficiency}
    \end{subfigure}\hfill
    \begin{subfigure}{0.48\linewidth}
        \centering
        \includegraphics[width=\linewidth]{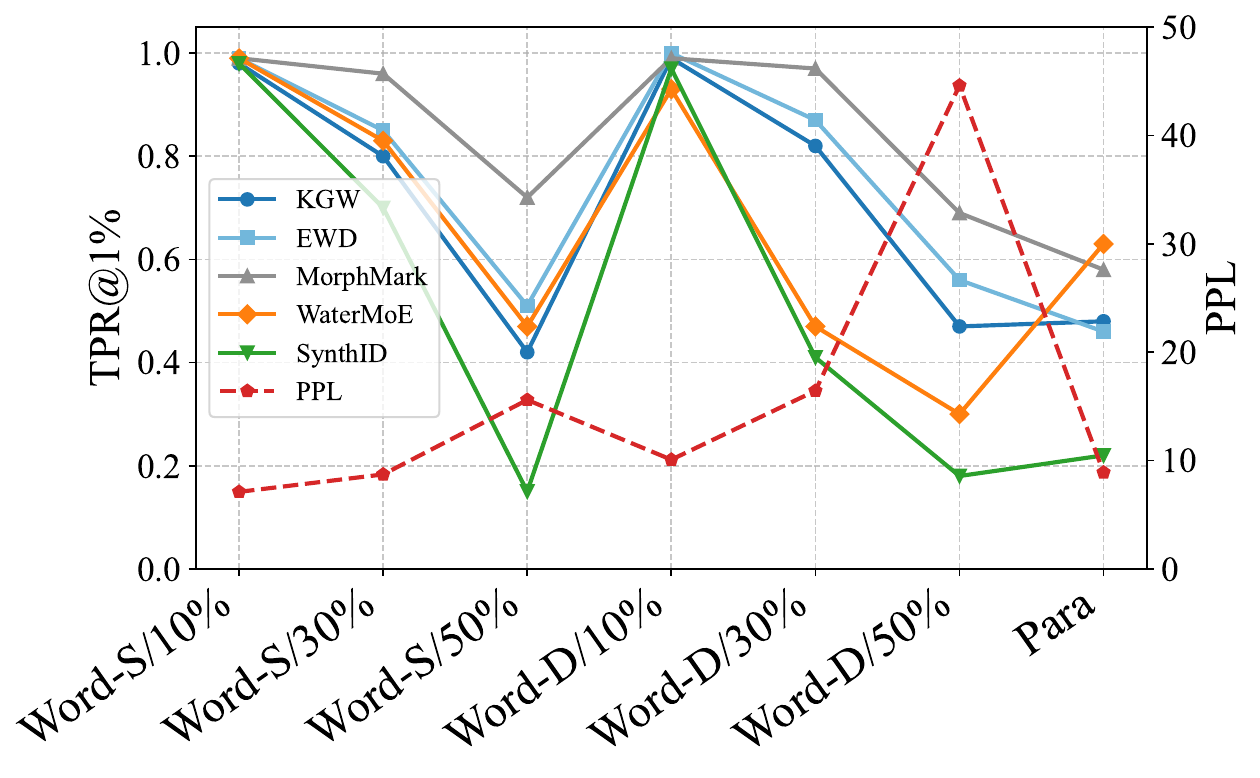}
        \caption{Robustness performance.}
        \label{fig:sub_robustness}
    \end{subfigure}
    
    \caption{
    (a) We report per-token insertion latency overhead (\%) and average detection time per sample (ms) on Mixtral-8x7B (C4) across different methods. 
    (b) Robustness performance of different methods, along with the corresponding perplexity (PPL), under word substitution, word deletion, and paraphrasing attacks on the Mixtral-8x7B over the C4 dataset. 
    }
    \label{fig:combined_eff_robust}
\end{figure}

\subsection{RQ3: Efficiency Analysis}

As shown in Fig.~\ref{fig:sub_efficiency}, WaterMoE introduces negligible computational overhead, with a clear advantage on the insertion side. WaterMoE achieves an insertion latency of only \textbf{1.1\%}, significantly lower than all baselines. This is enabled by the GPU-resident expert mapping design, which avoids extra data movement and integrates seamlessly into the inference pipeline. Our detection latency remains comparable to existing methods (around 150ms per sample), mostly due to the non-autoregressive scheme that performs a single prefill to obtain expert routing distributions for the entire sequence. In practice, insertion efficiency is more crucial as online generation is latency-critical, while detection occurs sparsely upon requests.

\subsection{RQ4: Robustness under Adversarial Attacks}

As shown in Fig.~\ref{fig:sub_robustness}, WaterMoE's robustness differ across attacks. It achieves competitive robustness under synonym substitution by wordnet, but shows a noticeable degradation under aggressive word deletion. This behavior stems from the design of WaterMoE, which relies on the preservation of local contextual coherence. When a large fraction of tokens is removed, the underlying linguistic structure is severely disrupted, weakening the statistical signal for detection. We note that such strong deletion-based perturbations, while commonly adopted in prior work, often lead to text that is syntactically unnatural or semantically degraded, reflected by a sharp increase in perplexity (PPL) from 7 to 45 in our setting. In many real-world scenarios, adversaries are more likely to apply semantics-preserving transformations rather than heavily destructive edits to maintain usability of the generated content. To better reflect this practical threat model, we use GLM-4.5 for full-text paraphrasing. The detection rate remains at 63\% TPR@1\% for WaterMoE, indicating moderate robustness under strong paraphrasing attacks. We additionally evaluate robustness under an API-assisted word-level substitution with $\sim30\%$ token replacement. WaterMoE maintains a 98\% TPR@1\% detection rate under the perturbation.

\subsection{RQ5: Stealthiness Analysis}
Beyond detectability by authorized parties, a desirable watermark should remain statistically indistinguishable from unwatermarked text under general-purpose analysis. We evaluate this property using two representative black-box statistical detectors~\cite{gloaguen2024black}, namely R-G detection and FIXED detection, which are designed to identify common watermarking artifacts, including distributional bias and diminished generation diversity. As shown in Table~\ref{tab:stealthiness}, existing methods are detectable under at least one setting, while WaterMoE consistently achieves high $p$-values across all cases. 
This indicates that its outputs remain statistically indistinguishable from the unwatermarked text, demonstrating strong stealthiness.

\begin{table}[t]
\centering
\small

\caption{Stealthiness evaluation measured by $p$-values of statistical watermark detectors. 
A $p$-value below $0.05$ indicates watermark detection.}

\begin{tabular}{lcccc}
\toprule
\multirow{2}{*}{Method} 
& \multicolumn{2}{c}{Mixtral-8$\times$7B} 
& \multicolumn{2}{c}{Qwen} \\
\cmidrule(lr){2-3} \cmidrule(lr){4-5}
& R-G Detection & FIXED Detection 
& R-G Detection & FIXED Detection \\
\midrule
Unwatermarked & 1.00 & 0.94 & 1.00 & 0.869 \\
KGW-series    & 0.00 & 0.94 & 0.00 & 0.94 \\
SynthID       & 0.06 & 0.94 & 0.04 & 0.94 \\
EXP           & 1.00 & $1.8\times10^{-6}$ & 1.00 & $1.5\times10^{-7}$ \\
\midrule
\textbf{WaterMoE}      & 0.92 & 0.94 & 1.00 & 0.94 \\
\bottomrule
\end{tabular}

\label{tab:stealthiness}
\end{table}

\subsection{Sensitivity to Routing Bias}

We analyze the sensitivity of the proposed watermarking method to the routing bias strength $\delta$ by varying it from 0.2 to 1.2. 
As shown in Table~\ref{tab:hyper_bias}, the detection performance remains consistently high across a wide range of bias values. 
In particular, both TPR and ROC-AUC exhibit near-saturated performance, while the perplexity (PPL) increases only marginally as the bias grows. 
This demonstrates that our method is robust to the choice of routing bias and can be easily applied in a plug-and-play manner without careful hyperparameter tuning.





\begin{wraptable}{r}{0.6\linewidth}
\vspace{-10pt}
\centering
\footnotesize
\caption{Sensitivity to routing bias on Mixtral-8x7B evaluated on the C4 dataset. We report detection performance (TPR@1, TPR@5, ROC-AUC) and generation quality (PPL).}
\begin{tabular}{c|cccc}
\hline
$\delta$ & TPR@1  & TPR@5 & ROC-AUC & PPL \\
\hline
0.2  & 0.96 & 0.98 & 0.99 & 6.59 \\
0.4  & 0.98 & 0.99 & 0.99 & 6.62 \\
0.6  & 0.98 & 0.99 & 0.99 & 6.62 \\
0.8  & 1.00 & 1.00 & 1.00 & 6.63 \\
1.0  & 1.00 & 1.00 & 1.00 & 6.70 \\
1.2  & 1.00 & 1.00 & 1.00 & 6.72 \\
\hline
\end{tabular}
\vspace{-10pt}
\label{tab:hyper_bias}
\end{wraptable}

\section{Limitation}
\label{Limitations}
We acknowledge that our current design is tailored to Mixture-of-Experts architectures. However, this choice is motivated by the increasing adoption of sparse activation paradigms in modern LLM systems (e.g., DeepSeek, Qwen-MoE), where routing naturally provides a structured and low-overhead carrier for watermark signals.
For dense Transformer models, a promising direction is to emulate implicit expert partitioning through structured sparsification. Specifically, one could leverage techniques such as dynamic neuron activation or structured pruning (e.g., block-wise or head-wise sparsity) to create multiple functionally redundant subspaces within dense layers.

\section{Conclusion}

Watermarking for LLMs remains challenging due to performance degradation and inference overhead, particularly in complex and constrained tasks. In this work, we systematically identify these limitations through a comprehensive benchmark and propose WaterMoE, a watermarking scheme tailored for MoE-based LLMs. By integrating watermark signals into the routing process, WaterMoE achieves strong detectability while preserving model quality and efficiency. Extensive experiments demonstrate its consistent advantages over existing methods, highlighting its practicality for real-world deployment.


\newpage
\bibliographystyle{unsrt}
\bibliography{reference}

@inproceedings{labruna2025retrieve,
  title={When to retrieve: Teaching llms to utilize information retrieval effectively},
  author={Labruna, Tiziano and Campos, Jon Ander and Azkune, Gorka},
  booktitle={Proceedings of the 15th International Conference on Recent Advances in Natural Language Processing-Natural Language Processing in the Generative AI Era},
  pages={623--632},
  year={2025}
}

@article{zhu2025large,
  title={Large language models for information retrieval: A survey},
  author={Zhu, Yutao and Yuan, Huaying and Wang, Shuting and Liu, Jiongnan and Liu, Wenhan and Deng, Chenlong and Chen, Haonan and Liu, Zheng and Dou, Zhicheng and Wen, Ji-Rong},
  journal={ACM Transactions on Information Systems},
  volume={44},
  number={1},
  pages={1--54},
  year={2025},
  publisher={ACM New York, NY}
}

@article{zeng2025rethinking,
  title={Rethinking Driving World Model as Synthetic Data Generator for Perception Tasks},
  author={Zeng, Kai and Wu, Zhanqian and Xiong, Kaixin and Wei, Xiaobao and Guo, Xiangyu and Zhu, Zhenxin and Ho, Kalok and Zhou, Lijun and Zeng, Bohan and Lu, Ming and others},
  journal={arXiv preprint arXiv:2510.19195},
  year={2025}
}

@inproceedings{ghodratnama2023adapting,
  title={Adapting llms for efficient, personalized information retrieval: Methods and implications},
  author={Ghodratnama, Samira and Zakershahrak, Mehrdad},
  booktitle={International conference on service-oriented computing},
  pages={17--26},
  year={2023},
  organization={Springer}
}

@article{wang2025survey,
  title={A survey of llm-based agents in medicine: How far are we from baymax?},
  author={Wang, Wenxuan and Ma, Zizhan and Wang, Zheng and Wu, Chenghan and Ji, Jiaming and Chen, Wenting and Li, Xiang and Yuan, Yixuan},
  journal={Findings of the Association for Computational Linguistics: ACL 2025},
  pages={10345--10359},
  year={2025}
}

@inproceedings{yuan2024continued,
  title={A continued pretrained llm approach for automatic medical note generation},
  author={Yuan, Dong and Rastogi, Eti and Naik, Gautam and Rajagopal, Sree Prasanna and Goyal, Sagar and Zhao, Fen and Chintagunta, Bharath and Ward, Jeffrey},
  booktitle={Proceedings of the 2024 Conference of the North American Chapter of the Association for Computational Linguistics: Human Language Technologies (Volume 2: Short Papers)},
  pages={565--571},
  year={2024}
}

@inproceedings{wang2023review,
  title={A review on code generation with llms: Application and evaluation},
  author={Wang, Jianxun and Chen, Yixiang},
  booktitle={2023 IEEE International Conference on Medical Artificial Intelligence (MedAI)},
  pages={284--289},
  year={2023},
  organization={IEEE}
}

@article{yang2025empirical,
  title={An empirical study of retrieval-augmented code generation: Challenges and opportunities},
  author={Yang, Zezhou and Chen, Sirong and Gao, Cuiyun and Li, Zhenhao and Hu, Xing and Liu, Kui and Xia, Xin},
  journal={ACM Transactions on Software Engineering and Methodology},
  volume={34},
  number={7},
  pages={1--28},
  year={2025},
  publisher={ACM New York, NY}
}

@article{jiang2026survey,
  title={A survey on large language models for code generation},
  author={Jiang, Juyong and Wang, Fan and Shen, Jiasi and Kim, Sungju and Kim, Sunghun},
  journal={ACM Transactions on Software Engineering and Methodology},
  volume={35},
  number={2},
  pages={1--72},
  year={2026},
  publisher={ACM New York, NY}
}

@article{huang2024understanding,
  title={Understanding the planning of llm agents: A survey},
  author={Huang, Xu and Liu, Weiwen and Chen, Xiaolong and Wang, Xingmei and Wang, Hao and Lian, Defu and Wang, Yasheng and Tang, Ruiming and Chen, Enhong},
  journal={arXiv preprint arXiv:2402.02716},
  year={2024}
}

@article{liu2023dynamic,
  title={Dynamic llm-agent network: An llm-agent collaboration framework with agent team optimization},
  author={Liu, Zijun and Zhang, Yanzhe and Li, Peng and Liu, Yang and Yang, Diyi},
  journal={arXiv preprint arXiv:2310.02170},
  year={2023}
}

@article{chu2025llm,
  title={Llm agents for education: Advances and applications},
  author={Chu, Zhendong and Wang, Shen and Xie, Jian and Zhu, Tinghui and Yan, Yibo and Ye, Jinheng and Zhong, Aoxiao and Hu, Xuming and Liang, Jing and Yu, Philip S and others},
  journal={arXiv preprint arXiv:2503.11733},
  volume={2},
  year={2025},
  publisher={Mar}
}

@inproceedings{li2026agentswift,
  title={Agentswift: Efficient llm agent design via value-guided hierarchical search},
  author={Li, Yu and Li, Lehui and Wu, Zhihao and Liao, Qingmin and Hao, Jianye and Shao, Kun and Xu, Fengli},
  booktitle={Proceedings of the AAAI Conference on Artificial Intelligence},
  volume={40},
  number={38},
  pages={31843--31851},
  year={2026}
}

@article{liang2026watermarking,
  title={Watermarking techniques for large language models: A survey},
  author={Liang, Yuqing and Xiao, Jiancheng and Gan, Wensheng and Yu, Philip S},
  journal={Artificial Intelligence Review},
  volume={59},
  number={2},
  pages={74},
  year={2026},
  publisher={Springer}
}

@article{huang2026rlspoofer,
  title={RLSpoofer: A Lightweight Evaluator for LLM Watermark Spoofing Resilience},
  author={Huang, Hanbo and Gong, Xuan and Zhang, Yiran and Zheng, Hao and Liang, Shiyu},
  journal={arXiv preprint arXiv:2604.11546},
  year={2026}
}

@inproceedings{kirchenbauer2023watermark,
  title={A watermark for large language models},
  author={Kirchenbauer, John and Geiping, Jonas and Wen, Yuxin and Katz, Jonathan and Miers, Ian and Goldstein, Tom},
  booktitle={International conference on machine learning},
  pages={17061--17084},
  year={2023},
  organization={PMLR}
}

@inproceedings{lu2024entropy,
  title={An entropy-based text watermarking detection method},
  author={Lu, Yijian and Liu, Aiwei and Yu, Dianzhi and Li, Jingjing and King, Irwin},
  booktitle={Proceedings of the 62nd Annual Meeting of the Association for Computational Linguistics (Volume 1: Long Papers)},
  pages={11724--11735},
  year={2024}
}

@article{kuditipudi2023robust,
  title={Robust distortion-free watermarks for language models},
  author={Kuditipudi, Rohith and Thickstun, John and Hashimoto, Tatsunori and Liang, Percy},
  journal={arXiv preprint arXiv:2307.15593},
  year={2023}
}

@article{hu2023unbiased,
  title={Unbiased watermark for large language models},
  author={Hu, Zhengmian and Chen, Lichang and Wu, Xidong and Wu, Yihan and Zhang, Hongyang and Huang, Heng},
  journal={arXiv preprint arXiv:2310.10669},
  year={2023}
}

@article{liu2023unforgeable,
  title={An unforgeable publicly verifiable watermark for large language models},
  author={Liu, Aiwei and Pan, Leyi and Hu, Xuming and Li, Shu'ang and Wen, Lijie and King, Irwin and Yu, Philip S},
  journal={arXiv preprint arXiv:2307.16230},
  year={2023}
}

@article{dathathri2024scalable,
  title={Scalable watermarking for identifying large language model outputs},
  author={Dathathri, Sumanth and See, Abigail and Ghaisas, Sumedh and Huang, Po-Sen and McAdam, Rob and Welbl, Johannes and Bachani, Vandana and Kaskasoli, Alex and Stanforth, Robert and Matejovicova, Tatiana and others},
  journal={Nature},
  volume={634},
  number={8035},
  pages={818--823},
  year={2024},
  publisher={Nature Publishing Group UK London}
}

@inproceedings{wang2025morphmark,
  title={Morphmark: Flexible adaptive watermarking for large language models},
  author={Wang, Zongqi and Gu, Tianle and Wu, Baoyuan and Yang, Yujiu},
  booktitle={Proceedings of the 63rd Annual Meeting of the Association for Computational Linguistics (Volume 1: Long Papers)},
  pages={4842--4860},
  year={2025}
}

@article{liu2023semantic,
  title={A semantic invariant robust watermark for large language models},
  author={Liu, Aiwei and Pan, Leyi and Hu, Xuming and Meng, Shiao and Wen, Lijie},
  journal={arXiv preprint arXiv:2310.06356},
  year={2023}
}

@inproceedings{he2024can,
  title={Can watermarks survive translation? on the cross-lingual consistency of text watermark for large language models},
  author={He, Zhiwei and Zhou, Binglin and Hao, Hongkun and Liu, Aiwei and Wang, Xing and Tu, Zhaopeng and Zhang, Zhuosheng and Wang, Rui},
  booktitle={Proceedings of the 62nd Annual Meeting of the Association for Computational Linguistics (Volume 1: Long Papers)},
  pages={4115--4129},
  year={2024}
}

@inproceedings{hou2024semstamp,
  title={Semstamp: A semantic watermark with paraphrastic robustness for text generation},
  author={Hou, Abe and Zhang, Jingyu and He, Tianxing and Wang, Yichen and Chuang, Yung-Sung and Wang, Hongwei and Shen, Lingfeng and Van Durme, Benjamin and Khashabi, Daniel and Tsvetkov, Yulia},
  booktitle={Proceedings of the 2024 Conference of the North American Chapter of the Association for Computational Linguistics: Human Language Technologies (Volume 1: Long Papers)},
  pages={4067--4082},
  year={2024}
}

@inproceedings{hou2024k,
  title={k-SemStamp: A clustering-based semantic watermark for detection of machine-generated text},
  author={Hou, Abe and Zhang, Jingyu and Wang, Yichen and Khashabi, Daniel and He, Tianxing},
  booktitle={Findings of the Association for Computational Linguistics: ACL 2024},
  pages={1706--1715},
  year={2024}
}

@article{cai2024survey,
  title={A survey on mixture of experts},
  author={Cai, Weilin and Jiang, Juyong and Wang, Fan and Tang, Jing and Kim, Sunghun and Huang, Jiayi},
  journal={Authorea Preprints},
  year={2024},
  publisher={Authorea}
}

@inproceedings{zhu2024llama,
  title={Llama-moe: Building mixture-of-experts from llama with continual pre-training},
  author={Zhu, Tong and Qu, Xiaoye and Dong, Daize and Ruan, Jiacheng and Tong, Jingqi and He, Conghui and Cheng, Yu},
  booktitle={Proceedings of the 2024 conference on empirical methods in natural language processing},
  pages={15913--15923},
  year={2024}
}

@inproceedings{rajbhandari2022deepspeed,
  title={Deepspeed-moe: Advancing mixture-of-experts inference and training to power next-generation ai scale},
  author={Rajbhandari, Samyam and Li, Conglong and Yao, Zhewei and Zhang, Minjia and Aminabadi, Reza Yazdani and Awan, Ammar Ahmad and Rasley, Jeff and He, Yuxiong},
  booktitle={International conference on machine learning},
  pages={18332--18346},
  year={2022},
  organization={PMLR}
}

@article{li2025uni,
  title={Uni-moe: Scaling unified multimodal llms with mixture of experts},
  author={Li, Yunxin and Jiang, Shenyuan and Hu, Baotian and Wang, Longyue and Zhong, Wanqi and Luo, Wenhan and Ma, Lin and Zhang, Min},
  journal={IEEE Transactions on Pattern Analysis and Machine Intelligence},
  year={2025},
  publisher={IEEE}
}

@inproceedings{pan2024markllm,
  title={Markllm: An open-source toolkit for llm watermarking},
  author={Pan, Leyi and Liu, Aiwei and He, Zhiwei and Gao, Zitian and Zhao, Xuandong and Lu, Yijian and Zhou, Binglin and Liu, Shuliang and Hu, Xuming and Wen, Lijie and others},
  booktitle={Proceedings of the 2024 Conference on Empirical Methods in Natural Language Processing: System Demonstrations},
  pages={61--71},
  year={2024}
}

@article{jiang2024mixtral,
  title={Mixtral of experts},
  author={Jiang, Albert Q and Sablayrolles, Alexandre and Roux, Antoine and Mensch, Arthur and Savary, Blanche and Bamford, Chris and Chaplot, Devendra Singh and Casas, Diego de las and Hanna, Emma Bou and Bressand, Florian and others},
  journal={arXiv preprint arXiv:2401.04088},
  year={2024}
}

@article{raffel2020exploring,
  title={Exploring the limits of transfer learning with a unified text-to-text transformer},
  author={Raffel, Colin and Shazeer, Noam and Roberts, Adam and Lee, Katherine and Narang, Sharan and Matena, Michael and Zhou, Yanqi and Li, Wei and Liu, Peter J},
  journal={Journal of machine learning research},
  volume={21},
  number={140},
  pages={1--67},
  year={2020}
}

@inproceedings{kryscinski2022booksum,
  title={Booksum: A collection of datasets for long-form narrative summarization},
  author={Kry{\'s}ci{\'n}ski, Wojciech and Rajani, Nazneen and Agarwal, Divyansh and Xiong, Caiming and Radev, Dragomir},
  booktitle={Findings of the association for computational linguistics: EMNLP 2022},
  pages={6536--6558},
  year={2022}
}

@inproceedings{fabbri2019multi,
  title={Multi-news: A large-scale multi-document summarization dataset and abstractive hierarchical model},
  author={Fabbri, Alexander Richard and Li, Irene and She, Tianwei and Li, Suyi and Radev, Dragomir},
  booktitle={Proceedings of the 57th annual meeting of the association for computational linguistics},
  pages={1074--1084},
  year={2019}
}

@inproceedings{fan2019eli5,
  title={ELI5: Long form question answering},
  author={Fan, Angela and Jernite, Yacine and Perez, Ethan and Grangier, David and Weston, Jason and Auli, Michael},
  booktitle={Proceedings of the 57th annual meeting of the association for computational linguistics},
  pages={3558--3567},
  year={2019}
}

@article{hendrycks2021measuring,
  title={Measuring coding challenge competence with apps},
  author={Hendrycks, Dan and Basart, Steven and Kadavath, Saurav and Mazeika, Mantas and Arora, Akul and Guo, Ethan and Burns, Collin and Puranik, Samir and He, Horace and Song, Dawn and others},
  journal={arXiv preprint arXiv:2105.09938},
  year={2021}
}

@article{li2022competition,
  title={Competition-level code generation with alphacode},
  author={Li, Yujia and Choi, David and Chung, Junyoung and Kushman, Nate and Schrittwieser, Julian and Leblond, R{\'e}mi and Eccles, Tom and Keeling, James and Gimeno, Felix and Dal Lago, Agustin and others},
  journal={Science},
  volume={378},
  number={6624},
  pages={1092--1097},
  year={2022},
  publisher={American Association for the Advancement of Science}
}

@article{cobbe2021training,
  title={Training verifiers to solve math word problems},
  author={Cobbe, Karl and Kosaraju, Vineet and Bavarian, Mohammad and Chen, Mark and Jun, Heewoo and Kaiser, Lukasz and Plappert, Matthias and Tworek, Jerry and Hilton, Jacob and Nakano, Reiichiro and others},
  journal={arXiv preprint arXiv:2110.14168},
  year={2021}
}

@article{hendrycks2020measuring,
  title={Measuring massive multitask language understanding},
  author={Hendrycks, Dan and Burns, Collin and Basart, Steven and Zou, Andy and Mazeika, Mantas and Song, Dawn and Steinhardt, Jacob},
  journal={arXiv preprint arXiv:2009.03300},
  year={2020}
}

@article{zhou2023instruction,
  title={Instruction-following evaluation for large language models},
  author={Zhou, Jeffrey and Lu, Tianjian and Mishra, Swaroop and Brahma, Siddhartha and Basu, Sujoy and Luan, Yi and Zhou, Denny and Hou, Le},
  journal={arXiv preprint arXiv:2311.07911},
  year={2023}
}

@article{wu2025writingbench,
  title={Writingbench: A comprehensive benchmark for generative writing},
  author={Wu, Yuning and Mei, Jiahao and Yan, Ming and Li, Chenliang and Lai, Shaopeng and Ren, Yuran and Wang, Zijia and Zhang, Ji and Wu, Mengyue and Jin, Qin and others},
  journal={arXiv preprint arXiv:2503.05244},
  year={2025}
}

@article{gloaguen2024black,
  title={Black-box detection of language model watermarks},
  author={Gloaguen, Thibaud and Jovanovi{\'c}, Nikola and Staab, Robin and Vechev, Martin},
  journal={arXiv preprint arXiv:2405.20777},
  year={2024}
}

@article{skliar2024mixture,
  title={Mixture of cache-conditional experts for efficient mobile device inference},
  author={Skliar, Andrii and van Rozendaal, Ties and Lepert, Romain and Boinovski, Todor and Van Baalen, Mart and Nagel, Markus and Whatmough, Paul and Bejnordi, Babak Ehteshami},
  journal={arXiv preprint arXiv:2412.00099},
  year={2024}
}

@article{yang2025qwen3,
  title={Qwen3 technical report},
  author={Yang, An and Li, Anfeng and Yang, Baosong and Zhang, Beichen and Hui, Binyuan and Zheng, Bo and Yu, Bowen and Gao, Chang and Huang, Chengen and Lv, Chenxu and others},
  journal={arXiv preprint arXiv:2505.09388},
  year={2025}
}

@article{liu2024deepseek,
  title={Deepseek-v3 technical report},
  author={Liu, Aixin and Feng, Bei and Xue, Bing and Wang, Bingxuan and Wu, Bochao and Lu, Chengda and Zhao, Chenggang and Deng, Chengqi and Zhang, Chenyu and Ruan, Chong and others},
  journal={arXiv preprint arXiv:2412.19437},
  year={2024}
}

\clearpage


\appendix

\section{Related Work}
\label{semantic_level}
\textbf{Semantic-level watermarking.}
Recent approaches explore embedding watermark signals at the semantic level. 
SIR~\cite{liu2023semantic} employs learned encoders to map semantically similar inputs to similar watermark patterns, improving robustness to paraphrasing. However, this design introduces additional system complexity and incurs non-trivial latency due to sentence-level encoding. Moreover, by conditioning watermark signals on semantic representations, such methods implicitly constrain the feasible token space during generation, leading to similar limitations as token-level approaches in restricting the solution space.

Subsequent methods such as SEMSTAMP~\cite{hou2024semstamp} and KSEMSTAMP~\cite{hou2024k} adopt a different strategy by discretizing semantic representations via locality-sensitive hashing or clustering (e.g., k-means), and enforcing constraints on the semantic space of the next sentence. In practice, these methods rely on repeated resampling until a candidate satisfying the watermark constraint is found. While effective in controlled settings, this mechanism introduces substantial computational overhead, often requiring tens of resampling iterations per sentence, which leads to significant latency.

More critically, the imposed semantic constraints further shrink the already limited feasible solution space in many real-world tasks, such as summarization and question answering, where correctness and coherence impose strict requirements. Empirical evidence shows that even with up to 50 resampling attempts, SEMSTAMP frequently fails to embed watermarks successfully on datasets such as MultiNews and ELI5 benchmarks. These limitations suggest that existing semantic-level watermarking methods are not well-suited for practical deployment in constraint-heavy generation scenarios.

\section{Detailed Background on Mixture-of-Experts (MoE)}
\label{appendix_moe}

\subsection{Overview}

Mixture-of-Experts (MoE) is a sparsely-activated neural architecture designed to scale model capacity without incurring proportional computational cost. Instead of using a single dense feed-forward network (FFN) in each Transformer layer, MoE replaces it with a set of $K$ parallel expert networks and a routing mechanism that dynamically selects a subset of experts for each token.

\begin{figure}[t]
\centering
\includegraphics[width=1\linewidth]{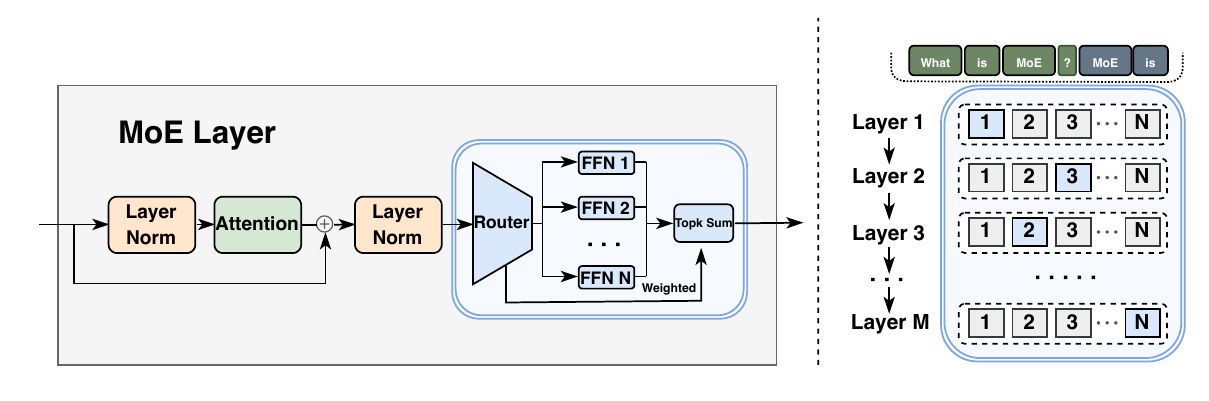}
\caption{Detailed Framework of MoE.}
\label{fig:moe_architecture}
\end{figure}

As illustrated in Fig.~\ref{fig:moe_architecture}, an MoE layer consists of three key components:
\begin{itemize}
    \item \textbf{Router (Gating Network):} computes routing scores for each expert.
    \item \textbf{Experts:} a set of independent feed-forward networks.
    \item \textbf{Aggregation:} combines outputs from selected experts via weighted summation.
\end{itemize}

This sparse activation enables MoE models to achieve extremely large parameter counts while keeping the per-token computation comparable to standard dense models.

\subsection{Formulation}

Given the hidden representation $\mathbf{h}_\ell \in \mathbb{R}^d$ at layer $\ell$, the router produces a score vector:
\begin{equation}
\mathbf{s}_\ell = \mathrm{Router}_\ell(\mathbf{h}_\ell), \quad \mathbf{s}_\ell \in \mathbb{R}^K,
\end{equation}
where $K$ is the number of experts.

These scores are transformed into a probability distribution via softmax:
\begin{equation}
p_\ell(i \mid \mathbf{h}_\ell) = \frac{\exp(s_{\ell,i})}{\sum_{j=1}^{K} \exp(s_{\ell,j})}.
\end{equation}

Instead of using all experts, only the top-$k$ experts are selected:
\begin{equation}
\mathcal{E}_\ell(\mathbf{h}_\ell) = \mathrm{TopK}(p_\ell(\cdot \mid \mathbf{h}_\ell), k).
\end{equation}

The final output of the MoE layer is:
\begin{equation}
\mathrm{MoE}_\ell(\mathbf{h}_\ell) = \sum_{i \in \mathcal{E}_\ell(\mathbf{h}_\ell)} p_\ell(i \mid \mathbf{h}_\ell) \cdot f_i(\mathbf{h}_\ell),
\end{equation}
where $f_i(\cdot)$ denotes the $i$-th expert network.

\subsection{Integration with Transformer}

In modern large language models, MoE layers replace the standard FFN sub-layer in Transformer blocks. A typical Transformer layer with MoE can be written as:
\begin{align}
\mathbf{h}'_\ell &= \mathbf{h}_\ell + \mathrm{Attention}(\mathrm{LN}(\mathbf{h}_\ell)), \\
\mathbf{h}_{\ell+1} &= \mathbf{h}'_\ell + \mathrm{MoE}_\ell(\mathrm{LN}(\mathbf{h}'_\ell)),
\end{align}
where $\mathrm{LN}(\cdot)$ denotes Layer Normalization.

\subsection{Routing Strategies}

Different routing strategies are used in practice:

\paragraph{Top-$k$ Routing}
The most common strategy selects the top-$k$ experts per token. Typical choices include:
\begin{itemize}
    \item $k=1$ (Switch Transformer): maximum efficiency
    \item $k \geq 2$ (GShard, DeepSeek): better performance with modest overhead
\end{itemize}

\subsection{Computational Efficiency}

Although the total number of parameters grows linearly with $K$, the actual computation per token depends only on $k$:
\begin{equation}
\text{FLOPs} \approx O(k \cdot d^2),
\end{equation}
which is comparable to a dense FFN when $k \ll K$.

Thus, MoE achieves:
\begin{itemize}
    \item \textbf{High capacity} (large parameter count)
    \item \textbf{Low compute cost} (sparse activation)
\end{itemize}

\section{Theoretical Analysis of Routing Loss under Biased Routing}
\label{appendix_prove}

In this section, we mathematically prove that the expected routing probability deviation caused by the watermark bias is strictly bounded by $\mathcal{O}(\delta^2)$. Thus, the routing process remains unbiased in expectation at the first order.

\subsection{Notations and Assumptions}

\noindent\textbf{Model State.} Let $\mathbf{s}_\ell \in \mathbb{R}^K$ denote the original routing logits at the $\ell$-th layer. The original routing probability for the $i$-th expert, obtained via the Softmax function, is defined as:
\begin{equation}
    p_i = \frac{e^{s_{\ell,i}}}{\sum_{j=1}^K e^{s_{\ell,j}}}.
\end{equation}

\noindent\textbf{Permutation-Symmetric Expert Map.} We define a binary indicator vector $\mathbf{g}_\ell \in \{0, 1\}^K$ to represent the green expert assignment at the current layer. Based on the expert map, the assignment is determined by $\mathcal{G}_\ell = \mathcal{M}_\ell(e_{\ell-1})$. This mapping satisfies two core constraints: (i) \textit{Fixed Green Proportion}, which dictates that exactly half of the experts are activated as green experts per routing, i.e., $\sum_{i=1}^K g_{\ell,i} = K/2$; and (ii) \textit{Permutation-Symmetric Constraint }, which ensures that in global expectation, every expert has an equal probability of being selected as a green expert:
\begin{equation}
    \Pr(g_{\ell,i} = 1) = \Pr(g_{\ell,i} = 0) = \frac{1}{2}, \quad \forall i \in \{1, \dots, K\}.
\end{equation}

\noindent\textbf{Bias Injection.} The actual routing bias vector applied at the current layer is $\mathbf{b}_\ell = \delta \mathbf{g}_\ell$. Due to the binary nature of $\mathbf{g}_\ell$, each component of the bias vector is restricted to a discrete set $b_{\ell,j} \in \{0, \delta\}$. We assume the bias magnitude $\delta$ is sufficiently small ($\delta \ll 1$) so that the higher-order Taylor expansion terms $\mathcal{O}(\|\mathbf{b}_\ell\|^3)$ can be reasonably truncated. The routing probability after bias injection is denoted as $p_i^{(g)}$.

\subsection{Proof of Zero First-Order Expectation}

According to the multidimensional Taylor expansion, the perturbed probability $p_i^{(g)}$ at $\mathbf{s}_\ell$ can be expanded as:
\begin{equation}
    p_i^{(g)} = p_i(\mathbf{s}_\ell) + \nabla p_i(\mathbf{s}_\ell)^\top \mathbf{b}_\ell + \frac{1}{2} \mathbf{b}_\ell^\top H^{(i)}(\mathbf{s}_\ell) \mathbf{b}_\ell + \mathcal{O}(\|\mathbf{b}_\ell\|^3).
\end{equation}
The first-order probability deviation is defined as $\Delta p_i^{(1)} = \nabla p_i(\mathbf{s}_\ell)^\top \mathbf{b}_\ell$. Expanding the inner product and taking the mathematical expectation $\mathbb{E}_{\mathbf{g}}$ over the bias distribution yields:
\begin{equation}
    \mathbb{E}_{\mathbf{g}}\left[ \Delta p_i^{(1)} \right] = \mathbb{E}_{\mathbf{g}}\left[ \sum_{j=1}^K \frac{\partial p_i}{\partial s_{\ell,j}} b_{\ell,j} \right] = \delta \sum_{j=1}^K \frac{\partial p_i}{\partial s_{\ell,j}} \mathbb{E}_{\mathbf{g}}[g_{\ell,j}].
\end{equation}
From the Permutation-Symmetric Constraint, we know $\mathbb{E}_{\mathbf{g}}[g_{\ell,j}] = \Pr(g_{\ell,j}=1) = 1/2$. Substituting this into the expectation gives:
\begin{equation}
    \mathbb{E}_{\mathbf{g}}\left[ \Delta p_i^{(1)} \right] = \frac{\delta}{2} \sum_{j=1}^K \frac{\partial p_i}{\partial s_{\ell,j}}.
\end{equation}
Using the partial derivative property of the Softmax function, we have:
\begin{equation}
\sum_{j=1}^K \frac{\partial p_i}{\partial s_{\ell,j}} = p_i(1 - p_i) - \sum_{j \neq i} p_i p_j = p_i - p_i \sum_{j=1}^K p_j. 
\end{equation}
Given the probability normalization property $\sum_{j=1}^K p_j = 1$, this sum strictly evaluates to 0. Therefore, the expected first-order deviation is completely eliminated, yielding $\mathbb{E}_{\mathbf{g}}[\Delta p_i^{(1)}] = 0$.

\subsection{Bounding the Second-Order Term}

The second-order error term is $\Delta p_i^{(2)} = \frac{1}{2} \mathbf{b}_\ell^\top H^{(i)}(\mathbf{s}_\ell) \mathbf{b}_\ell$. The quadratic form can be explicitly expressed as:
\begin{equation}
    \mathbf{b}_\ell^\top H^{(i)} \mathbf{b}_\ell = p_i \left[ (b_{\ell,i} - \bar{b})^2 - \left( \sum_{j=1}^K p_j b_{\ell,j}^2 - \bar{b}^2 \right) \right],
\end{equation}
where $\bar{b} = \sum_{j=1}^K p_j b_{\ell,j}$ represents the expected value of the bias vector under the original probability distribution $\mathbf{p}$. We now bound the two components inside the brackets.

\noindent\textbf{First term.} Since $b_{\ell,i} \in \{0, \delta\}$ and the weighted average satisfies $0 \le \bar{b} \le \delta$, the squared difference between them is strictly bounded by $(b_{\ell,i} - \bar{b})^2 \le \delta^2$.

\noindent\textbf{Second term.} This component represents the variance of the bias under distribution $\mathbf{p}$. Since the indicator variable is strictly binary ($g_{\ell,j} \in \{0, 1\}$), it inherently satisfies $g_{\ell,j}^2 = g_{\ell,j}$. For the bias terms $b_{\ell,j} = \delta g_{\ell,j}$, this implies:
\begin{equation}
    b_{\ell,j}^2 = (\delta g_{\ell,j})^2 = \delta^2 g_{\ell,j} = \delta (\delta g_{\ell,j}) = \delta b_{\ell,j}.
\end{equation}
Substituting this identity into the probability-weighted sum yields:
\begin{equation}
    \sum_{j=1}^K p_j b_{\ell,j}^2 = \sum_{j=1}^K p_j (\delta b_{\ell,j}) = \delta \left( \sum_{j=1}^K p_j b_{\ell,j} \right) = \delta \bar{b}.
\end{equation}
The entire second term then simplifies to a downward-opening quadratic function with respect to $\bar{b}$. By completing the square, we obtain:
\begin{equation}
    \sum_{j=1}^K p_j b_{\ell,j}^2 - \bar{b}^2 = \delta \bar{b} - \bar{b}^2 = -\left(\bar{b} - \frac{\delta}{2}\right)^2 + \frac{\delta^2}{4}.
\end{equation}
Because the squared component is non-negative, the global maximum of this term is achieved at $\bar{b} = \delta/2$, leading to the strict bound:
\begin{equation}
    \sum_{j=1}^K p_j b_{\ell,j}^2 - \bar{b}^2 \le \frac{\delta^2}{4}.
\end{equation}

Combining the bounds for both the first and second terms, the absolute value of the total second-order deviation is limited by:
\begin{equation}
    \left| \Delta p_i^{(2)} \right| \le \frac{1}{2} p_i \left( \delta^2 + \frac{\delta^2}{4} \right) = \frac{5}{8} p_i \delta^2 = \mathcal{O}(\delta^2).
\end{equation}

\subsection*{Conclusion}

Combining the derivations above, under the first-order approximation, the expected routing probability satisfies:
\begin{equation}
    \mathbb{E}_{\mathbf{g}}[p_i^{(g)}] = p_i + \mathbb{E}_{\mathbf{g}}[\Delta p_i^{(1)}] + \mathbb{E}_{\mathbf{g}}[\Delta p_i^{(2)}] = p_i + 0 + \mathcal{O}(\delta^2).
\end{equation}
This confirms that the watermark injection preserves the expected routing probabilities (i.e., $\mathbb{E}_{\mathbf{g}}[p_i^{(g)}] = p_i$ at the first order), demonstrating that the routing process remains unbiased in expectation.

\section{Performance on Moderate-Complexity Tasks.}
\label{appendix_modcom}

Table~\ref{tab:moderate_results} shows the detection-text quality trade-off on both MultiNews and ELI5. On the smaller Mixtral-8$\times$7B model, which is harder to achieve balance, WaterMoE achieves substantially higher detection rates under strict thresholds while also attaining the highest AUC and maintaining Rouge-L comparable to the unwatermarked baseline. A similar trend holds on ELI5, where WaterMoE achieves the best AUC and Rouge-L, with strong detection performance. Overall, it achieves strong detectability without sacrificing text quality across models and tasks.

\begin{table}[htbp]
  \caption{Watermarking performance on Moderate-Complexity Tasks. WaterMoE consistently achieves the highest AUC and TPR while maintaining competitive Rouge-L scores.}
  \label{tab:moderate_results}
  \centering
  \resizebox{\linewidth}{!}{
  \begin{tabular}{lcccccccc}
    \toprule
    
    \multicolumn{9}{c}{\textbf{Dataset: MultiNews}} \\
    \midrule
    \multirow{2}{*}{\textbf{Algorithm}} & \multicolumn{4}{c}{\textbf{Mixtral-8$\times$7B}} & \multicolumn{4}{c}{\textbf{Qwen}} \\
    \cmidrule(lr){2-5} \cmidrule(lr){6-9}
    & \textbf{TPR@1\%} & \textbf{TPR@5\%} & \textbf{AUC} & \textbf{Rouge-L} & \textbf{TPR@1\%} & \textbf{TPR@5\%} & \textbf{AUC} & \textbf{Rouge-L} \\
    \midrule
    
    \rowcolor{gray!15} \textit{Unwatermarked} & - & - & - & 38.1 & - & - & - & 40.6 \\
    EWD \cite{lu2024entropy}& 71.0 & 84.5 & 94.2 & 37.8 & 91.5 & 95.5 & 99.3 & 40.1 \\
    EXPEdit \cite{kuditipudi2023robust} & 41.5 & 61.5 & 91.7 & 35.3 & 36.0 & 48.0 & 77.6 & 31.5 \\
    KGW \cite{kirchenbauer2023watermark} & 21.0 & 49.5 & 86.5 & 36.3 & 98.5 & 100.0 & 100.0 & 40.3 \\
    MorphMark \cite{wang2025morphmark} & 26.0 & 41.0 & 88.2 & 36.4 & 77.0 & 93.5 & 98.9 & 40.5 \\
    SIR \cite{liu2023semantic} & 21.5 & 36.0 & 85.0 & 35.9 & - & - & - & - \\
    SynthID \cite{dathathri2024scalable} & 8.0 & 17.5 & 63.7 & 38.0 & 98.0 & 99.0 & 99.5 & \textbf{40.5} \\
    Unbiased \cite{hu2023unbiased} & 74.5 & 82.5 & 93.1 & 37.7 & 13.0 & 52.0 & 73.8 & 40.5 \\
    UPV  \cite{liu2023unforgeable}& 17.0 & 29.5 & 87.4 & 34.2 & 3.5 & 27.0 & 84.3 & 40.3 \\
    XSIR \cite{he2024can} & 8.0 & 19.5 & 67.5 & 35.3 & - & - & - & - \\
    \midrule
    \rowcolor{gray!15} \textbf{WaterMoE} & \textbf{83.5} & \textbf{86.5} & \textbf{99.6} & \textbf{38.3} & \textbf{100.0} & \textbf{100.0} & \textbf{100.0} & 40.4 \\
    
    \midrule
    \midrule
    
    \multicolumn{9}{c}{\textbf{Dataset: ELI5}} \\
    \midrule
    \multirow{2}{*}{\textbf{Algorithm}} & \multicolumn{4}{c}{\textbf{Mixtral-8$\times$7B}} & \multicolumn{4}{c}{\textbf{Qwen}} \\
    \cmidrule(lr){2-5} \cmidrule(lr){6-9}
    & \textbf{TPR@1\%} & \textbf{TPR@5\%} & \textbf{AUC} & \textbf{Rouge-L} & \textbf{TPR@1\%} & \textbf{TPR@5\%} & \textbf{AUC} & \textbf{Rouge-L} \\
    \midrule
    
    \rowcolor{gray!15} \textit{Unwatermarked} & - & - & - & 31.2 & - & - & - & 28.8 \\
    EWD & \textbf{87.0} & 91.0 & 98.6 & 29.3 & 100.0 & 100.0 & 100.0 & 28.5 \\
    EXPEdit & 28.5 & 28.5 & 64.9 & 29.4 & 32.0 & 32.0 & 69.8 & 26.1 \\
    KGW & 78.0 & 89.5 & 97.6 & 28.8 & 100.0 & 100.0 & 100.0 & 28.5 \\
    MorphMark & 49.5 & 64.5 & 90.8 & 29.3 & 92.0 & 96.5 & 99.5 & 28.5 \\
    SIR & 63.0 & 75.5 & 93.7 & 28.6 & - & - & - & - \\
    SynthID & 43.0 & 67.0 & 92.1 & 30.2 & 99.5 & 100.0 & 100.0 & 28.5 \\
    Unbiased & 84.0 & 89.5 & 96.4 & 29.8 & 100.0 & 100.0 & 100.0 & 28.0 \\
    UPV & 60.5 & 80.5 & 95.1 & 28.4 & 95.5 & 100.0 & 99.9 & 28.2 \\
    XSIR & 47.0 & 60.5 & 87.5 & 27.6 & - & - & - & - \\
    \midrule
    \rowcolor{gray!15} \textbf{WaterMoE} & 81.0 & \textbf{97.5} & \textbf{98.8} & \textbf{30.5} & \textbf{100.0} & \textbf{100.0} & \textbf{100.0} & \textbf{28.8} \\
    
    \bottomrule
  \end{tabular}
  }
\end{table}

\section{Three-Ent and Semantic Entropy Analysis}

\begin{figure}[t]
\centering
\includegraphics[width=0.7\linewidth]{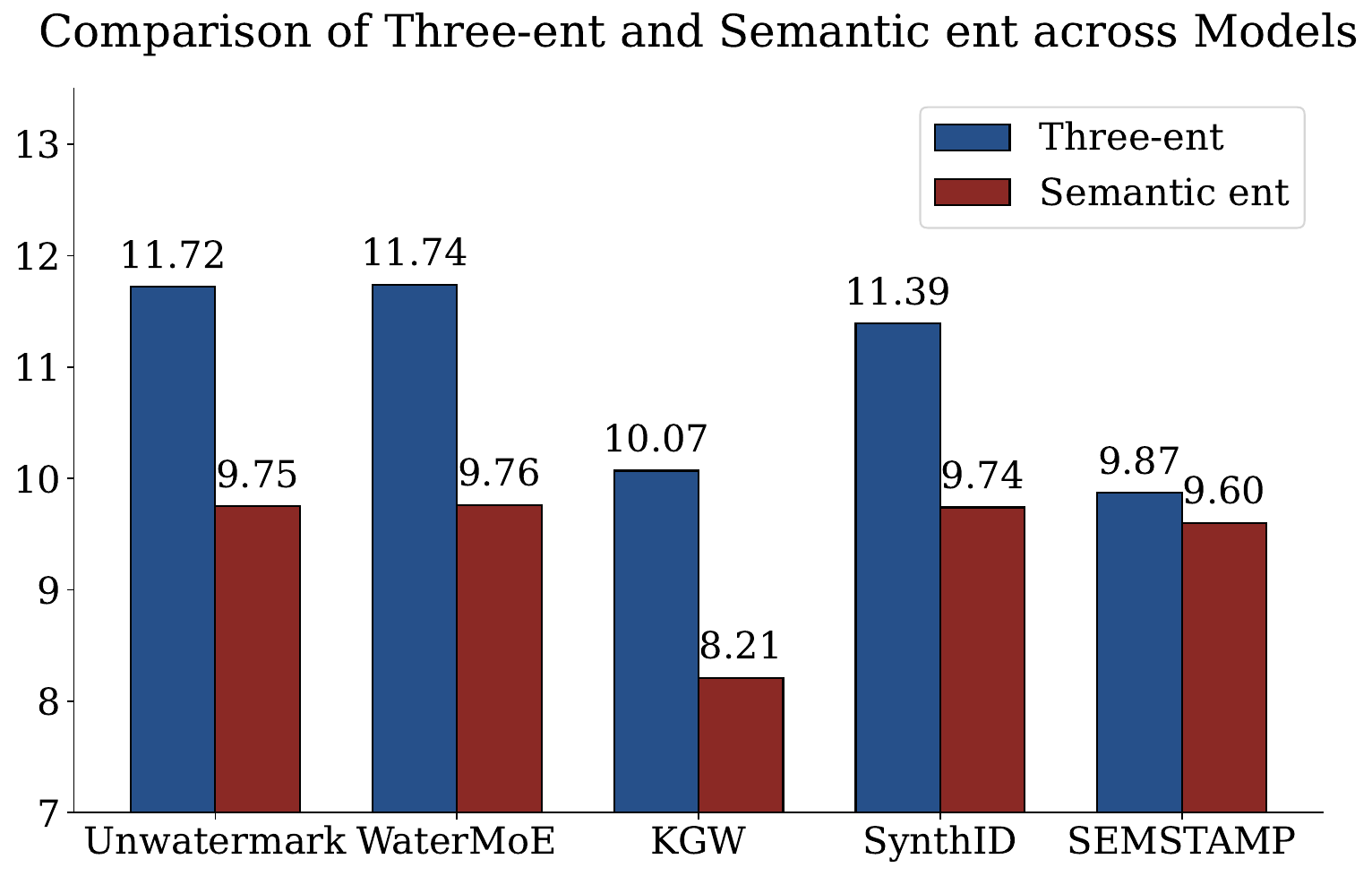}
\caption{Trigram entropy and semantic entropy comparison. WaterMoE preserves entropy closest to the non-watermarked model, indicating better text diversity.}
\label{fig:ent}
\end{figure}

As illustrated in Fig.~\ref{fig:ent}, we compare the trigram entropy (three-ent) and semantic entropy of texts generated by different methods. 
Our proposed WaterMoE produces entropy values that are closest to those of the non-watermarked model. 
In contrast, other watermarking methods lead to noticeable reductions in entropy. 

This observation indicates that WaterMoE better preserves the diversity and richness of the generated text, maintaining natural generation characteristics while embedding the watermark.

\section{Watermarking Setup for Reproducibility}
\label{appendix_watermark_setup}

To ensure reproducibility, we summarize the watermark-specific hyperparameters for all algorithms under different task complexity levels in Table~\ref{tab:wm_all}. Each cell reports the key hyperparameter used for the corresponding setting.

\begin{table*}[t]
\centering
\caption{Watermark hyperparameters across different task complexity levels. EXPEdit and Unbiased are distortion-free algorithms and thus do not require injection strength parameters (denoted by -). SIR and XSIR are not applicable to high-complexity tasks as they are incompatible with the Qwen model.}
\label{tab:wm_all}
\begin{tabular}{lccc}
\toprule
\textbf{Algorithm} 
& \textbf{Low} 
& \textbf{Moderate} 
& \textbf{High} \\
\midrule

KGW        
& $\delta = 2$ 
& $\delta = 2$ 
& $\delta = 5$ \\

SynthID    
& \texttt{num\_leaves=2} 
& \texttt{num\_leaves=2} 
& \texttt{num\_leaves=8} \\

EWD        
& $\delta = 2$ 
& $\delta = 2$ 
& $\delta = 5$ \\

EXPEdit    
& - 
& - 
& - \\

MorphMark  
& $k_{\text{exp}} = 1.3$ 
& $k_{\text{exp}} = 1.3$ 
& $k_{\text{exp}} = 7$ \\

Unbiased   
& - 
& - 
& - \\

UPV        
& $\delta = 2$ 
& $\delta = 2$ 
& $\delta = 4$ \\

SIR        
& $\delta = 1$ 
& $\delta = 1$ 
& - \\

XSIR       
& $\delta = 1$ 
& $\delta = 1$ 
& - \\

WaterMoE   
& $\delta = 0.8$ 
& $\delta = 0.2$ 
& $\delta = 0.2$ \\

\bottomrule
\end{tabular}
\end{table*}

\section{Dataset and Global Prompts}
\label{appendix:dataset}

To ensure evaluation consistency and standardize model behavior across diverse tasks, we define a specific "Global Prompt" for each dataset. These prompts serve as the foundational instructions that dictate the model's reasoning style and output constraints, effectively minimizing performance variance caused by prompt phrasing. 

\begin{figure}[htbp]
    \centering
    \includegraphics[width=\textwidth]{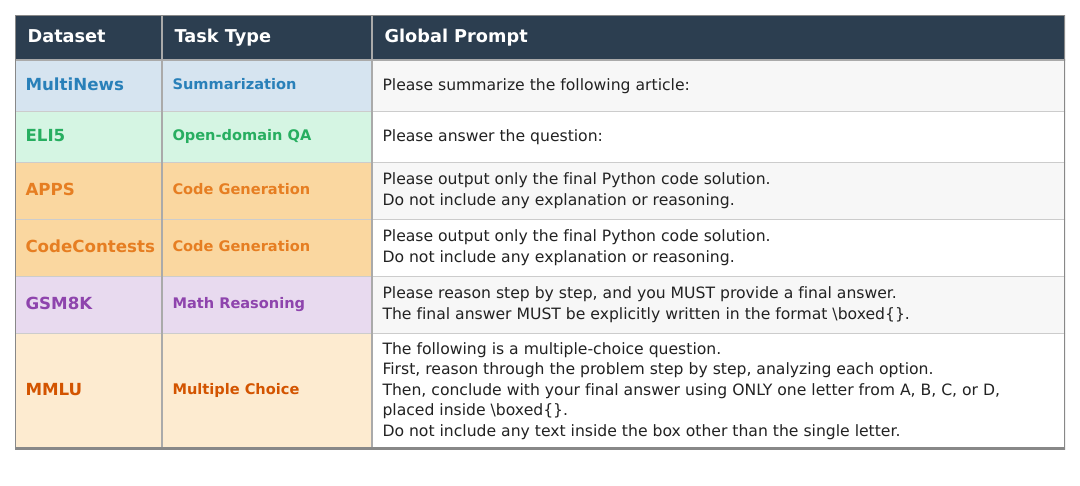}
    \caption{Detailed Global Prompts used across various datasets to standardize model instructions and output formats.}
    \label{fig:global_prompts}
\end{figure}

The complete set of these standardized instructions, categorized by task type and dataset, is presented in Figure \ref{fig:global_prompts}. These prompts are prepended to each individual query to guide the model's generation process throughout our experiments.

\section{Configuration of Attack Prompts}
\label{appendix:attack_prompts}

In this section, we present the prompts used for different types of watermark attacks. In addition to the document-level paraphrasing attack, we specifically design token-level substitution. 

\textbf{Token-level Substitution Attack.}
The prompt for the token-level substitution attack explicitly instructs the model to replace a specific ratio of words with their synonyms while maintaining the original semantic meaning and logical order as much as possible, as shown in Figure \ref{fig:prompt_substitution_attack_en}.

\begin{figure}[htbp]
    \centering
    \begin{tcolorbox}[colback=gray!10, colframe=black, boxrule=0.5pt, arc=4pt, width=0.95\textwidth]
    Please rewrite the following text by substituting words. \\
    \textbf{Requirement:} Replace exactly \{ratio\}\% of the words while keeping the overall semantics, sentence order, and logic unchanged. \\
    \textbf{Strategy:} \\
    The primary goal is to achieve the \{ratio\}\% substitution rate through synonym replacement. If a high substitution rate is requested, prioritize reaching the target ratio even if semantic coherence cannot be fully guaranteed. If a low substitution rate is requested, ensure that \{100-ratio\}\% of the text remains strictly unchanged. Output only the rewritten text without any additional explanation.\\
    \\
    \{text\}
    \end{tcolorbox}
    \caption{Prompt used for the token-level substitution attack.}
    \label{fig:prompt_substitution_attack_en}
\end{figure}

\textbf{Document-level Paraphrasing Attack.}
For the document-level paraphrasing attack, we instruct the model to rewrite the entire text output, simulating a real-world scenario of semantic restructuring, as depicted in Figure \ref{fig:prompt_paraphrasing_attack}.

\begin{figure}[htbp]
    \centering
    \begin{tcolorbox}[colback=gray!10, colframe=black, boxrule=0.5pt, arc=4pt, width=0.85\textwidth]
    Please rewrite the following text (Only return the rewritten text): \\
    \\
    \{text\}
    \end{tcolorbox}
    \caption{Prompt used for the document-level paraphrasing attack.}
    \label{fig:prompt_paraphrasing_attack}
\end{figure}

\section{ROC-AUC Curves of Watermarking Methods on MultiNews.}

\begin{figure*}[t]
\centering

\begin{subfigure}{0.32\textwidth}
    \centering
    \includegraphics[width=\linewidth]{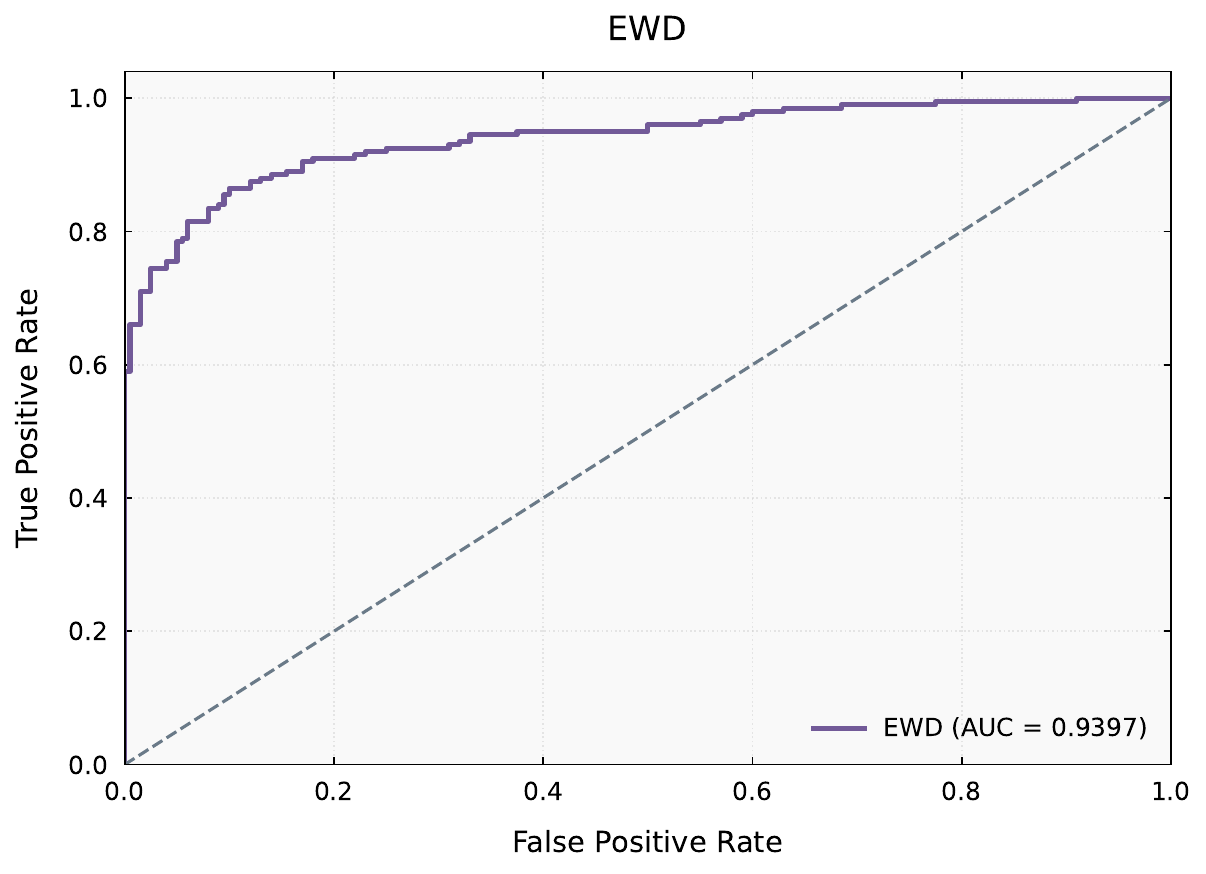}
    \caption{EWD}
\end{subfigure}
\hfill
\begin{subfigure}{0.32\textwidth}
    \centering
    \includegraphics[width=\linewidth]{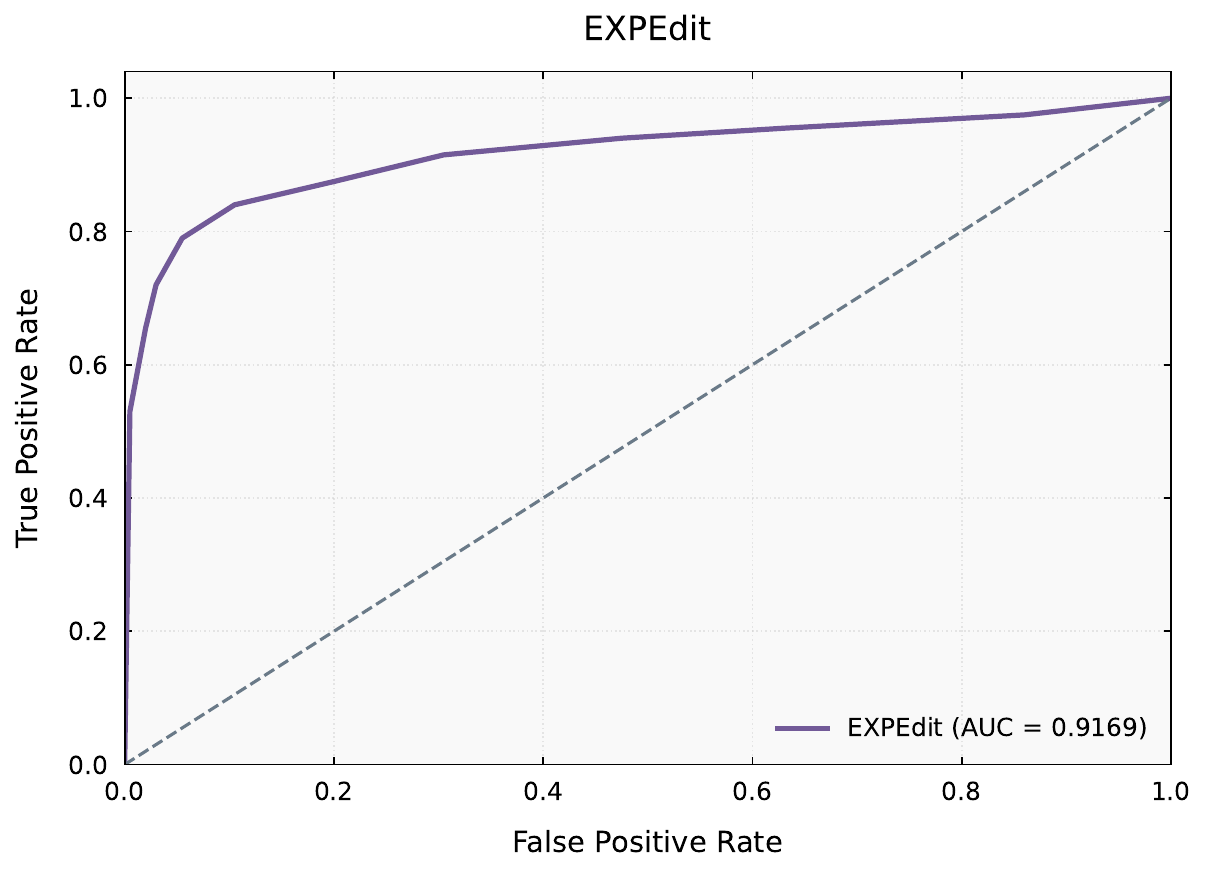}
    \caption{EXPEdit}
\end{subfigure}
\hfill
\begin{subfigure}{0.32\textwidth}
    \centering
    \includegraphics[width=\linewidth]{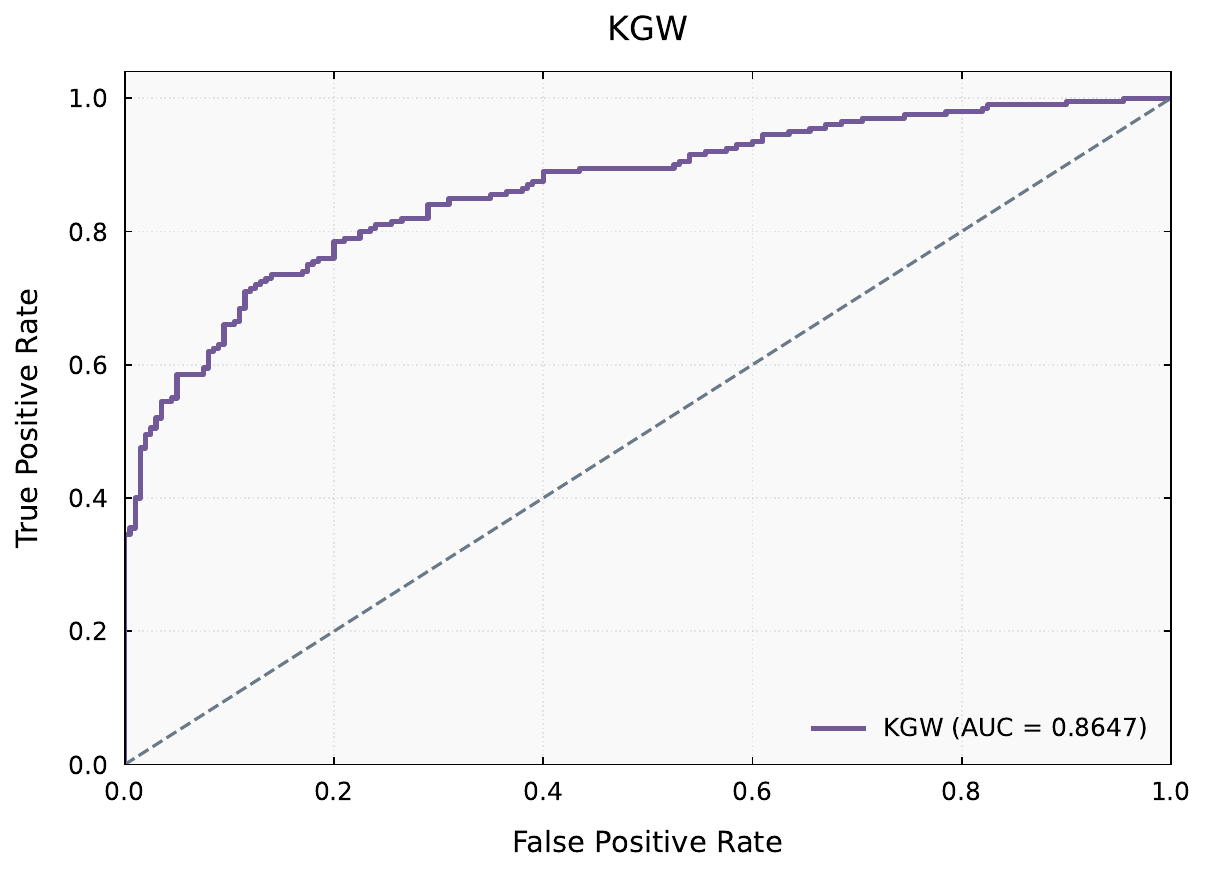}
    \caption{KGW}
\end{subfigure}

\vspace{0.5em}

\begin{subfigure}{0.32\textwidth}
    \centering
    \includegraphics[width=\linewidth]{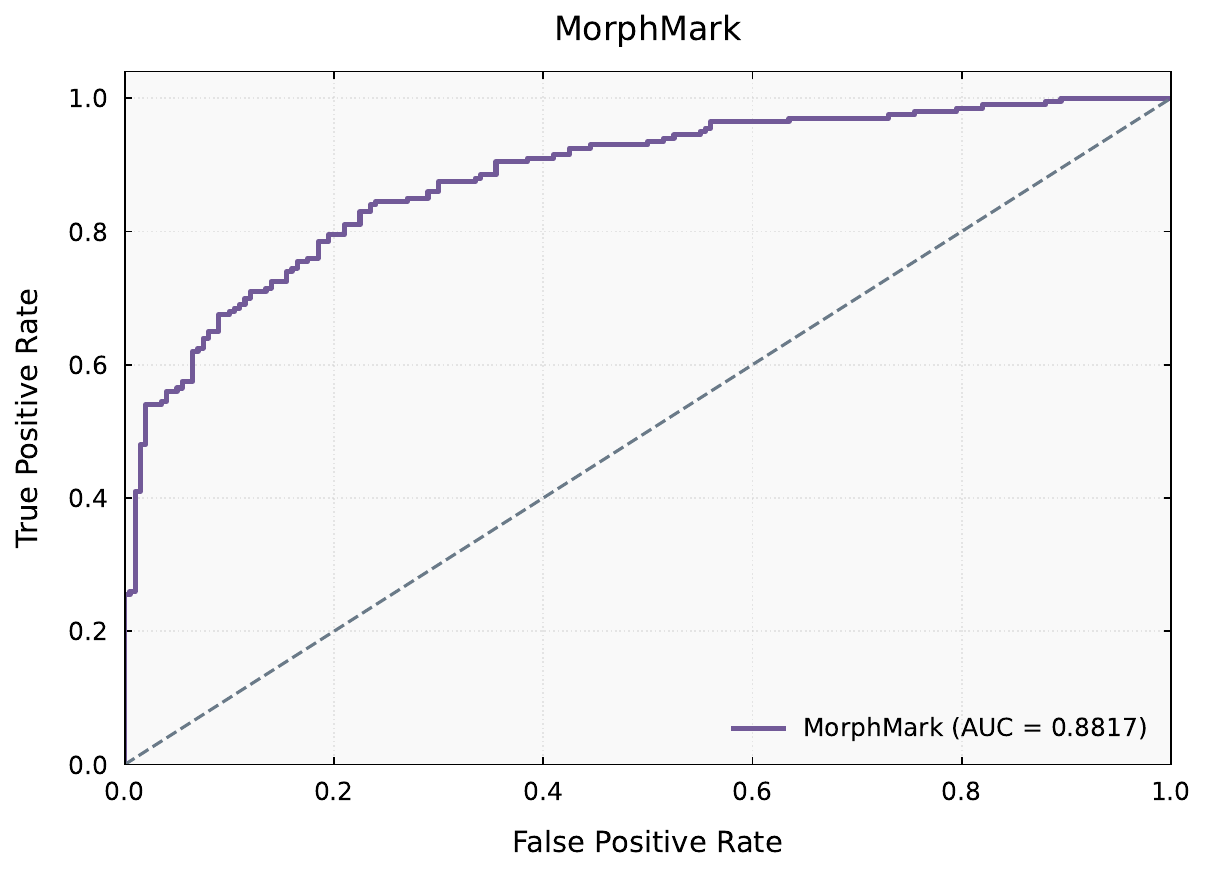}
    \caption{MorphMark}
\end{subfigure}
\hfill
\begin{subfigure}{0.32\textwidth}
    \centering
    \includegraphics[width=\linewidth]{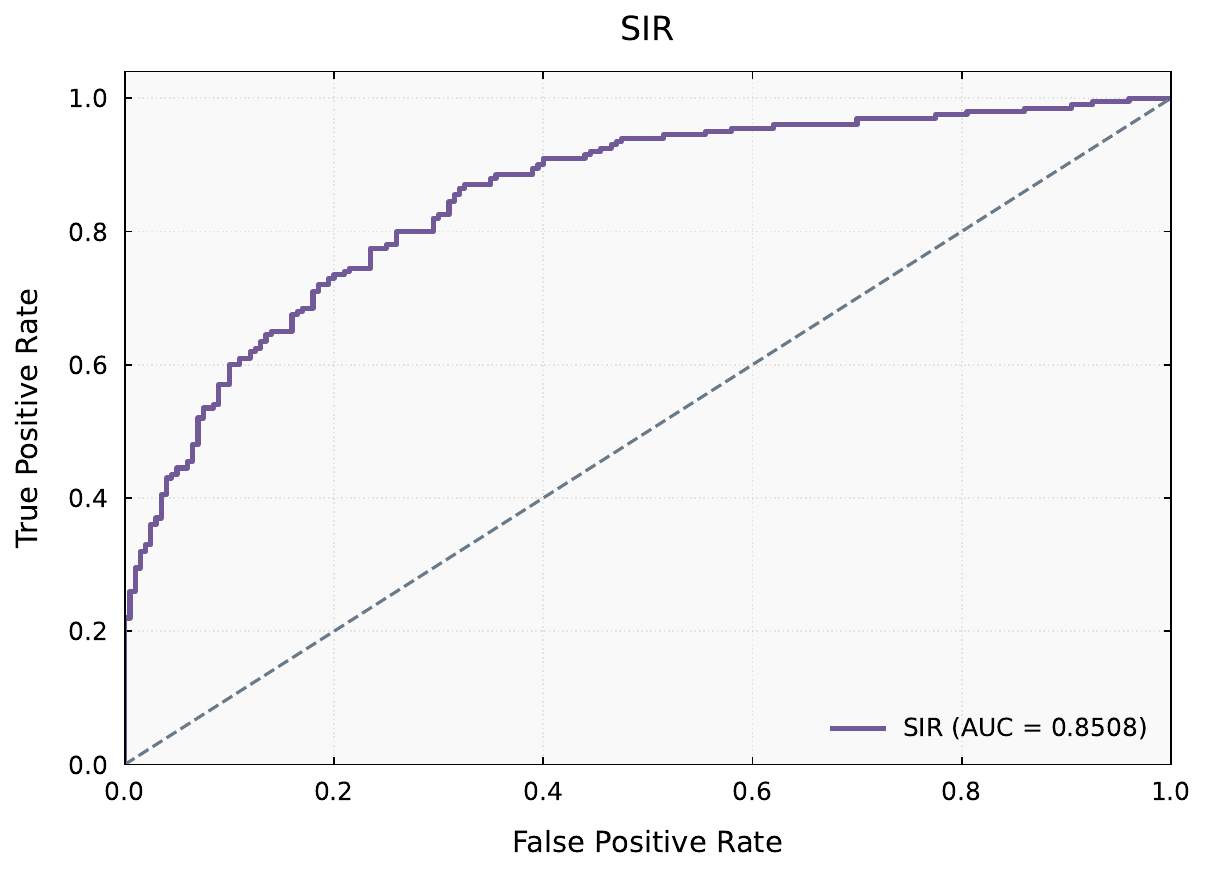}
    \caption{SIR}
\end{subfigure}
\hfill
\begin{subfigure}{0.32\textwidth}
    \centering
    \includegraphics[width=\linewidth]{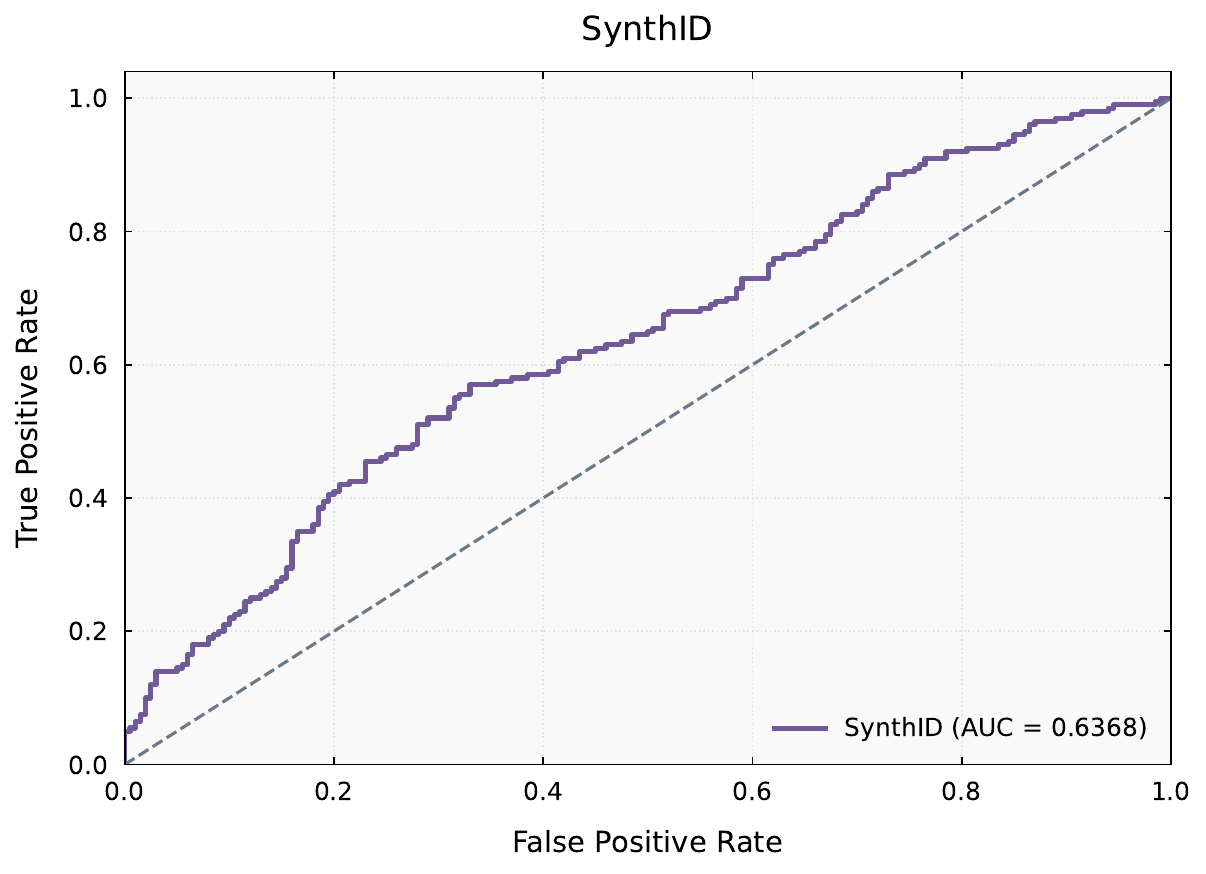}
    \caption{SynthID}
\end{subfigure}

\vspace{0.5em}

\begin{subfigure}{0.32\textwidth}
    \centering
    \includegraphics[width=\linewidth]{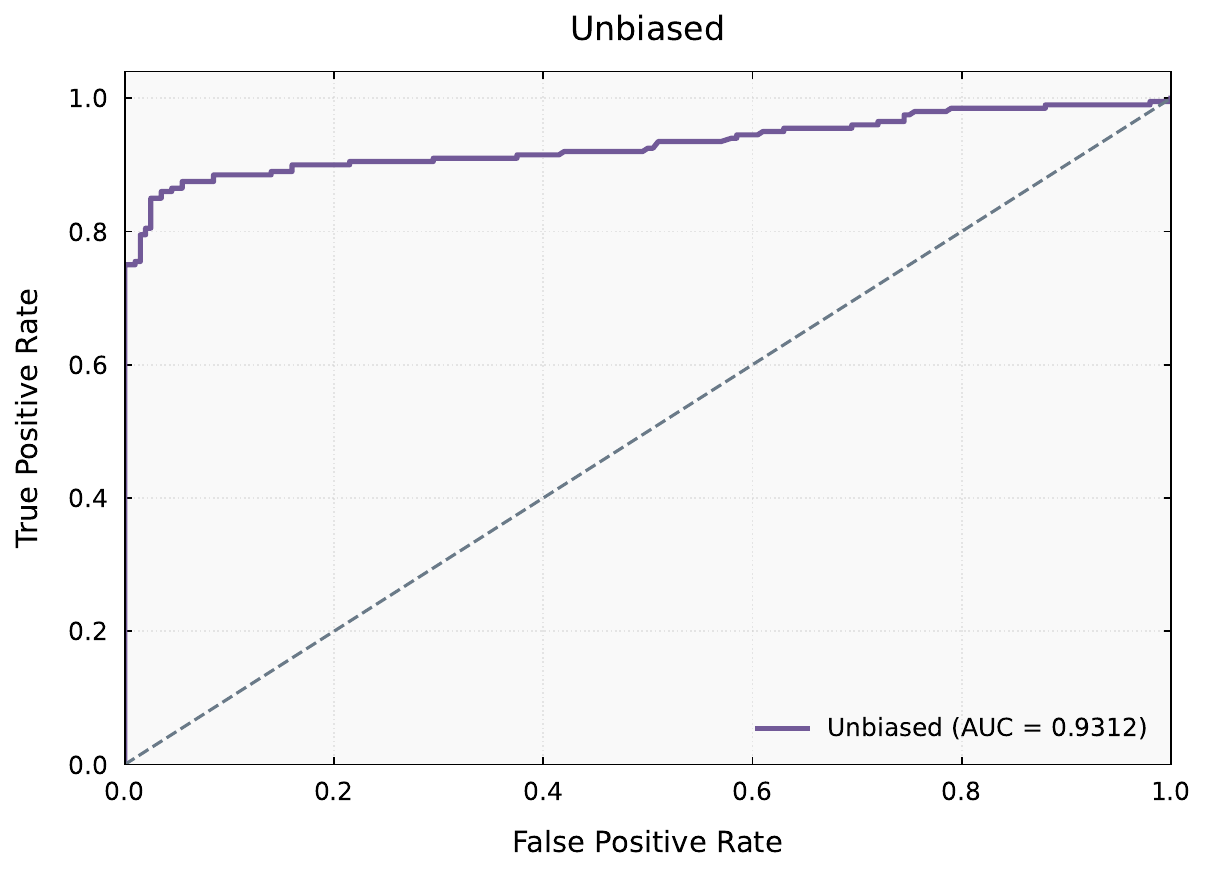}
    \caption{Unbiased}
\end{subfigure}
\hfill
\begin{subfigure}{0.32\textwidth}
    \centering
    \includegraphics[width=\linewidth]{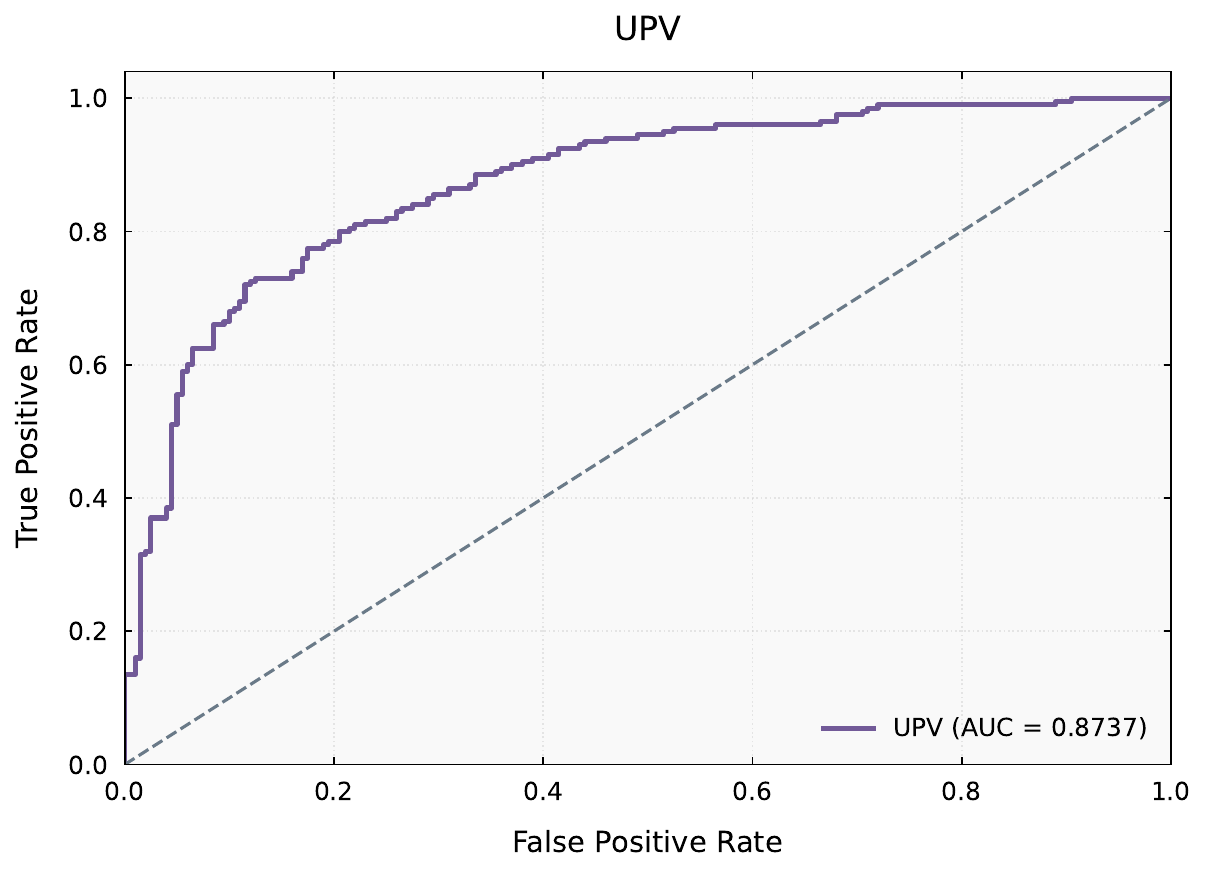}
    \caption{UPV}
\end{subfigure}
\hfill
\begin{subfigure}{0.32\textwidth}
    \centering
    \includegraphics[width=\linewidth]{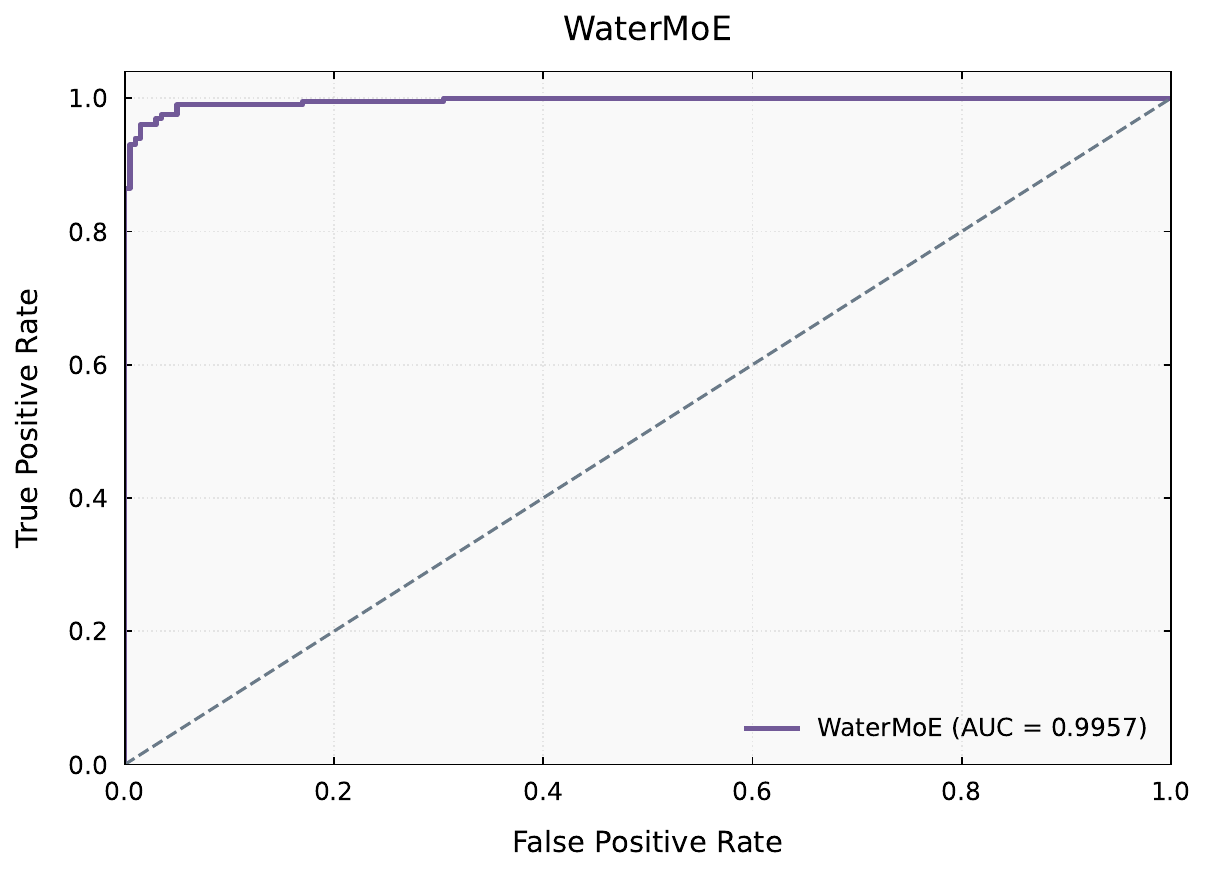}
    \caption{WaterMoE}
\end{subfigure}

\caption{ROC-AUC curves of different watermarking methods on MultiNews (Mixtral-8x7B).}
\label{fig:roc_multinews}
\end{figure*}

As shown in Fig.\ref{fig:roc_multinews}, we present the ROC-AUC curves of all evaluated watermarking algorithms on the MultiNews task using Mixtral-8x7B.

\section{Examples of Prompts and Generated Texts across Different Tasks}
\label{appendix:prompts_examples}

To provide a qualitative understanding of watermark behavior across different task types, we present representative examples from three categories: summarization, code generation, and alignment. For each task, we show the input prompt together with the corresponding outputs generated with and without watermarking. These examples highlight how WaterMoE embeds watermark signals while preserving task fidelity. In particular, the comparisons illustrate that the watermarked outputs remain semantically consistent with the unwatermarked ones, even under strict task constraints such as structured code generation and format-sensitive alignment tasks.
\begin{figure}[htbp]
    \centering
    \includegraphics[width=\textwidth]{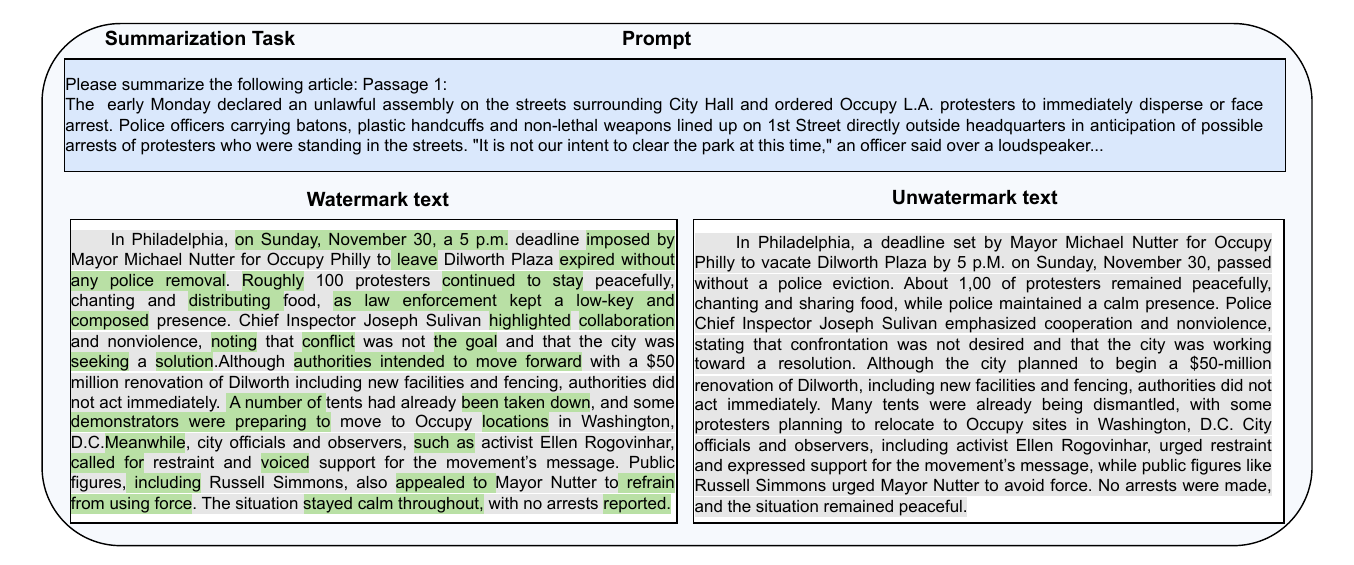}
    \caption{An example of the Summarization task from the Multi-News dataset. The texts are generated using the WaterMoE watermark algorithm. In the ``Watermark text'' section, words with a green background indicate tokens that were altered by the watermark algorithm, while words with a gray background indicate unchanged tokens that remain identical to those in the ``Unwatermark text''.}
    \label{fig:prompt-sum-task}
\end{figure}

\begin{figure}[htbp]
    \centering
    \includegraphics[width=\textwidth]{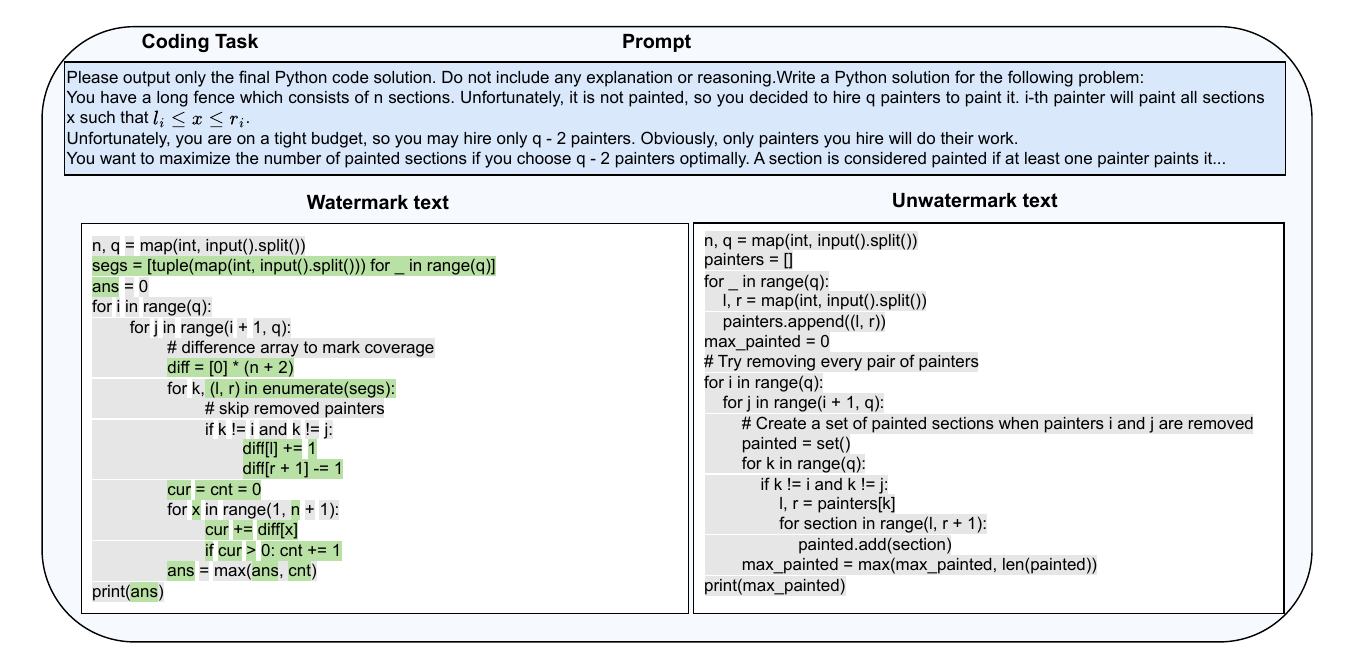}
    \caption{An example of the Coding task from the APPS dataset. The text demonstrates the prompt and the corresponding outputs generated with and without the WaterMoE watermark algorithm.}
    \label{fig:prompt-coding-task}
\end{figure}

\begin{figure}[htbp]
    \centering
    \includegraphics[width=\textwidth]{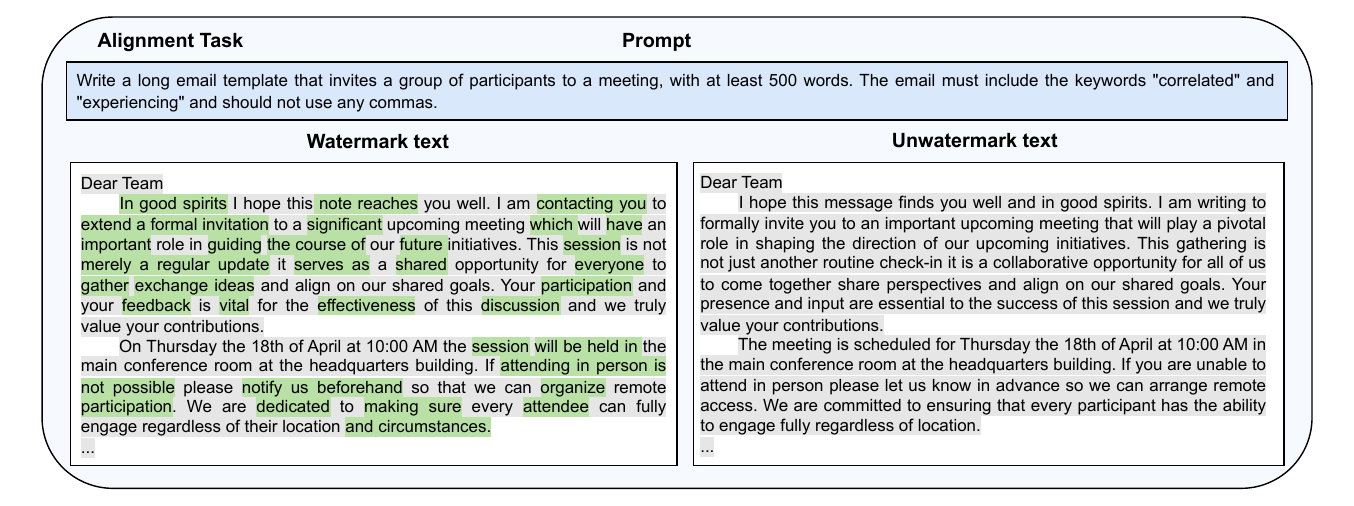}
    \caption{An example of the Alignment task from the IFEval dataset. The text illustrates the prompt along with the watermarked (using WaterMoE) and unwatermarked generation results, demonstrating adherence to the length and formatting constraints.}
    \label{fig:prompt-alignment-task}
\end{figure}

\clearpage

\end{document}